\documentclass[Crown]{sagej}

\usepackage{moreverb,url}

\usepackage{amsmath,amssymb,amsthm,amsfonts}
\usepackage{mathabx}
\usepackage{mathrsfs}
\usepackage{bm}
\usepackage{graphicx}
\usepackage{color,xcolor}
\usepackage{tabularx}
\usepackage[caption=false]{subfig}
\usepackage{epsfig,psfrag}
\usepackage{tikz}
\usepackage{pgflibraryshapes}
\usepackage{graphicx,psfrag}
\usepackage{apptools}
\usepackage[square,numbers,sort&compress]{natbib}
\AtAppendix{\counterwithin{theorem}{section}}

\newcommand{\be}{\begin{equation}}
\newcommand{\en}{\end{equation}}
\newcommand{\la}{\label}

\newcommand{\paa}{\partial}
\def\rr#1{(\ref{#1})}
\def\bm#1{\mbox{\boldmath{$#1$}}}
\def\ii{{\rm i}}

\def \theequation {\thesection.\arabic{equation}}

\numberwithin{equation}{section}
\theoremstyle{plain}

\newtheorem{theorem*}{Theorem}

\theoremstyle{definition}

\newcommand{\A}{\mathcal{A}}

\DeclareMathOperator{\tr}{tr}
\DeclareMathOperator{\Div}{Div}
\DeclareMathOperator{\ddiv}{div}

\DeclareMathOperator{\diag}{diag}

\makeatletter
\newcommand{\varnabla}{{\mathpalette\a@nabla\relax}}
\newcommand\a@nabla[2]{%
	\setbox\z@=\hbox{$\m@th#1\bigtriangledown$}%
	\ht\z@.7\ht\z@
	\raise\dp\z@\box\z@
}
\makeatother

\usepackage[colorlinks,bookmarksopen,bookmarksnumbered,citecolor=red,urlcolor=red]{hyperref}

\newcommand\BibTeX{{\rmfamily B\kern-.05em \textsc{i\kern-.025em b}\kern-.08em
T\kern-.1667em\lower.7ex\hbox{E}\kern-.125emX}}

\def\volumeyear{2016}

\begin{document}

\runninghead{Xiang Yu and Yibin Fu}

\title{On the incremental equations in surface elasticity}

\author{Xiang Yu\affilnum{1} and Yibin Fu\affilnum{2}}

\affiliation{\affilnum{1}Department of Mathematics, School of Computer Science and Technology, Dongguan University of Technology, Dongguan, 523808,  China\\
\affilnum{2}School of Computer Science and Mathematics, Keele University, Staffs ST5 5BG, UK}

\corrauth{Xiang Yu}

\email{yuxiang@dgut.edu.cn}

\begin{abstract}
We derive the incremental equations for a hyperelastic solid that incorporate surface tension effect by assuming that the surface energy is a general function of the surface deformation gradient. The incremental equations take the same simple form as their purely mechanical counterparts and are valid for any geometry. In particular, for isotropic materials, the extra surface elastic moduli are expressed in terms of the surface energy function and the two surface principal stretches. The effectiveness of the resulting incremental theory is illustrated by applying it to study the Plateau--Rayleigh and Wilkes instabilities in a solid cylinder.
\end{abstract}


\maketitle

\section{Introduction}
Professor Ray Ogden's published work has always been marked by simplicity, accessibility, and elegance in its presentations.
For instance, for small deformations superimposed on a finite deformation in the purely mechanical setting, he promoted the use of the incremental equations of motion in its most simple form \citep{chadwick1971definition}
\begin{align}
{\cal A}_{jilk} u_{k,lj}=\rho \ddot{u}_i,   \la{eqn1}
\end{align}
and for isotropic hyperelastic materials  gave the following compact expressions for the elastic moduli \citep{ogden1984non}:
\begin{align}
\begin{split}\la{eqn3}
J{\cal A}_{iijj}=&\lambda_i\lambda_jW_{ij},   \\
J{\cal A}_{ijkl}=&\frac{\lambda_i^2}{\lambda_i^2-\lambda_j^2}
(\lambda_iW_i-\lambda_jW_j)\delta_{ik}\delta_{jl}  \\ &
+\frac{\lambda_i\lambda_j}{\lambda_i^2-\lambda_j^2}
(\lambda_jW_i-\lambda_iW_j)\delta_{il}\delta_{jk}, \quad
i\ne j,
\end{split}
\end{align}
where $(u_i)$ is the incremental displacement,  $\rho$ is the material density in the finitely deformed configuration, a superimposed dot denotes differentiation with respect to time, ${\cal A}_{jilk}$ are the first-order instantaneous elastic moduli, $W$ is the strain energy that is viewed as a function of the three principal stretches $\lambda_1, \lambda_2$ and $\lambda_3$, $J=\lambda_1\lambda_2\lambda_3$, and $W_i=\paa W/\paa\lambda_i,\,
W_{ij}=\paa^2W/\paa\lambda_i\paa\lambda_j$, etc.
Although presented above for compressible isotropic materials and in terms of rectangular coordinates, the above formulation can easily be adapted for incompressible materials and more general coordinates, and the nonlinear version of \rr{eqn1} can also be obtained \citep{fu1999nonlinear}. The elegance of this formulation has also been extended to electroelasticity \citep{DO2014ijes}, magnetoelasticity \citep{DO2011MMS, SO2011JAM}, and materials with initial stresses \citep{DO2021JEM, MOD2021IJES}. Although we nowadays take these equations for granted, it is with these elegant and most accessible formulations that the dynamic and stability properties of various problems have been studied on a firm footing and in a systematic manner. In particular, we note that equation \rr{eqn1} takes the same form as those in general anisotropic elasticity; all the information about the finite deformation is nicely hidden in the elastic moduli. As a result, all the advanced methods developed for anisotropic materials, such as the Stroh formalism \citep{stroh1958, stroh1962}, can be applied to finitely deformed elastic solids; see, e.g., \cite{fu-mielke2002} and \cite{sb2018}.

Our current study is motivated by the observation that a general incremental formulation that takes into account surface tension effect does not seem to exist and this has, to some extent, hampered effective studies and even given rise to controversies. Although Ogden, Steigmann and collaborators \citep{ogden1997effect,dryburgh1999bifurcation,ogden2002plane} have previously derived the incremental equations based on the Steigmann--Ogden theory \citep{steigmann1997plane}, their derivation is restricted to the plane-strain case. Due to the lack of a general incremental theory, previous studies on surface tension-induced instabilities have resorted  to rather {\it ad hoc}, although ingenious, approaches for specific geometries. For instance, recent studies by Taffetani and Ciarletta \cite{taffetani2015beading}, Xuan and Biggins \cite{xuan2017plateau}, Wang \cite{wang2020axisymmetric}, Fu et al. \cite{fu2021necking}, Emery and Fu \cite{emery2021localised,emery2021post}, and Emery \cite{emery2023elasto} have all considered the cylindrical geometry for which the governing equations are derived directly from a variational principle, with incompressibility automatically satisfied by the use of mixed coordinates and introduction of a stream function \cite{ciarletta2011generating}, and then linearised in order to conduct the necessary linear stability analysis. Bakiler et al. \cite{bakiler2023surface} derived the incremental equations for a compressible elastic cylinder but focused on a specific surface energy function.
Also, Gurtin and Murdoch's original theory \citep{ gurtin1975continuum, gurtin1978surface} is essentially an {\it ad hoc} incremental theory since it is not derived from a fully nonlinear variational principle and the finite deformation was not fully accounted for. The {\it ad hoc} nature of this theory has given way to different interpretations; see \cite{ru2010simple} for a detailed discussion. Such different interpretations have even given rise to controversies. For instance, recent results presented by Yang et al. \cite{yang2022surface} and Ru \cite{ru2022critical} are at variance with those given by Mora et al. \cite{mora2010capillarity}, and Taffetani and Ciarletta \cite{taffetani2015beading}.

Surface tension typically appears in boundary value problems through the non-dimensional parameter $\gamma/(\mu \ell)$, where $\gamma$ is a measure of the surface tension, $\mu$ the shear modulus, and $\ell$  a representative lengthscale in the problem (e.g., for a solid cylinder $\ell$ would be the radius). This parameter becomes significant when either the material is soft (small $\mu$) and/or $\ell$ becomes small. As a result, surface tension effect becomes non-negligible for soft materials at micrometer and sub-micrometer levels (e.g., nerve fibers), and has become an area of active research in recent decades \citep{liu2012elastocapillarity, javili2013thermomechanics}. Extensive research work \citep{cammarata1994surface,sharma2002interfacial,sharma2003effect,sharma2004size,duan2005size,duan2009theory,he2008surface,chhapadia2011curvature,gao2017curvature} has been devoted to studying the size-dependent elastic properties of nanomaterials induced by surface tension. Although surface tension in solid materials has been briefly discussed in early studies by Young \cite{young1832essay}, Laplace \cite{marquis1825traite},  Gibbs \cite{gibbs1906scientific}, Shuttleworth \cite{shuttleworth1950surface}, Scriven \cite{scriven1960dynamics}, and Orowan \cite{orowan1970surface}, it was not a proper topic of continuum mechanics until Gurtin and Murdoch published their pioneering works  \cite{gurtin1975continuum,gurtin1978surface}. However, the theory formulated by Gurtin and Murdoch is essentially a linearized incremental theory and was designed to address small deformations at small scales.  It was much later that fully nonlinear theories taking into account surface elasticity were developed; see  \cite{steigmann1997plane, steigmann1999elastic, huang2006theory,steinmann2008boundary,gao2014curvature,huang2020}.

The rest of this paper is divided into six sections as follows. After reviewing briefly the kinematics of deformable surfaces in Section \ref{sec:2}, we summarize and adapt slightly in Section \ref{sec:set} the surface elasticity theory given by Steinmann \cite{steinmann2008boundary}, presenting it in a self-contained manner. In Section \ref{sec:incremental}, we first derive the incremental governing equations in their general nonlinear forms and then obtain their linearized forms.  In Section \ref{sec:5}, we show how our incremental theory can be reduced to the Gurtin--Murdoch theory under an appropriate assumption. The effectiveness of our incremental theory is verified by considering the Plateau--Rayleigh and Wilkes instabilities in Section \ref{sec:6}. The paper is concluded in the final section with a summary and additional comments.

For convenience, we  present here some notation needed in the sequel. Let $\bm{A}$ and $\bm{B}$ be two tensors. Their double dot product is defined by $\bm{A}:\bm{B}=A_{ij}B_{ij}$. Given a scalar-valued function $W(\bm{A})$ of a tensor variable $\bm{A}$, its derivative is defined by $\Big(\frac{\partial W}{\partial \bm{A}}\Big)_{ij}=\frac{\partial W}{\partial A_{ij}}$. The summation convection over repeated indices is adopted and a comma preceding indices means differentiation. In a summation,  Greek letters $\alpha,\beta,\dots$
run from $1$ to $2$, whereas Latin letters $i, j,\cdots $run from $1$ to $3$.

\section{Kinematics of deformable surfaces}\label{sec:2}

Consider a hyperelastic solid that occupies a region $\Omega\subset\mathbb{R}^3$ in the reference configuration. The boundary of the region $\Omega$ is denoted by $\partial \Omega$ and is assumed to be piecewise smooth. The position of a material point in the reference configuration is denoted by $\bm{X}$. The solid deforms under the combined effect of mechanical forces and surface stresses. After deformation, the material point  $\bm{X}$ moves to a new position $\bm{x}$ under the deformation map
\begin{align}\label{eq:x}
\bm{x}=\phi(\bm{X}),\quad \bm{X}\in \Omega.
\end{align}
We assume that $\phi$ is at least twice continuously differentiable on $\Omega\cup\partial \Omega$. The  deformation gradient $\bm{F}$ is defined by 
\begin{align}
\bm{F}=\frac{\partial\bm{x}}{\partial\bm{X}},
\end{align}
which means $d\bm{x}= \bm{F} d\bm{X}$.

\begin{figure}[htbp]
	\centering
	\includegraphics[width=0.9\linewidth]{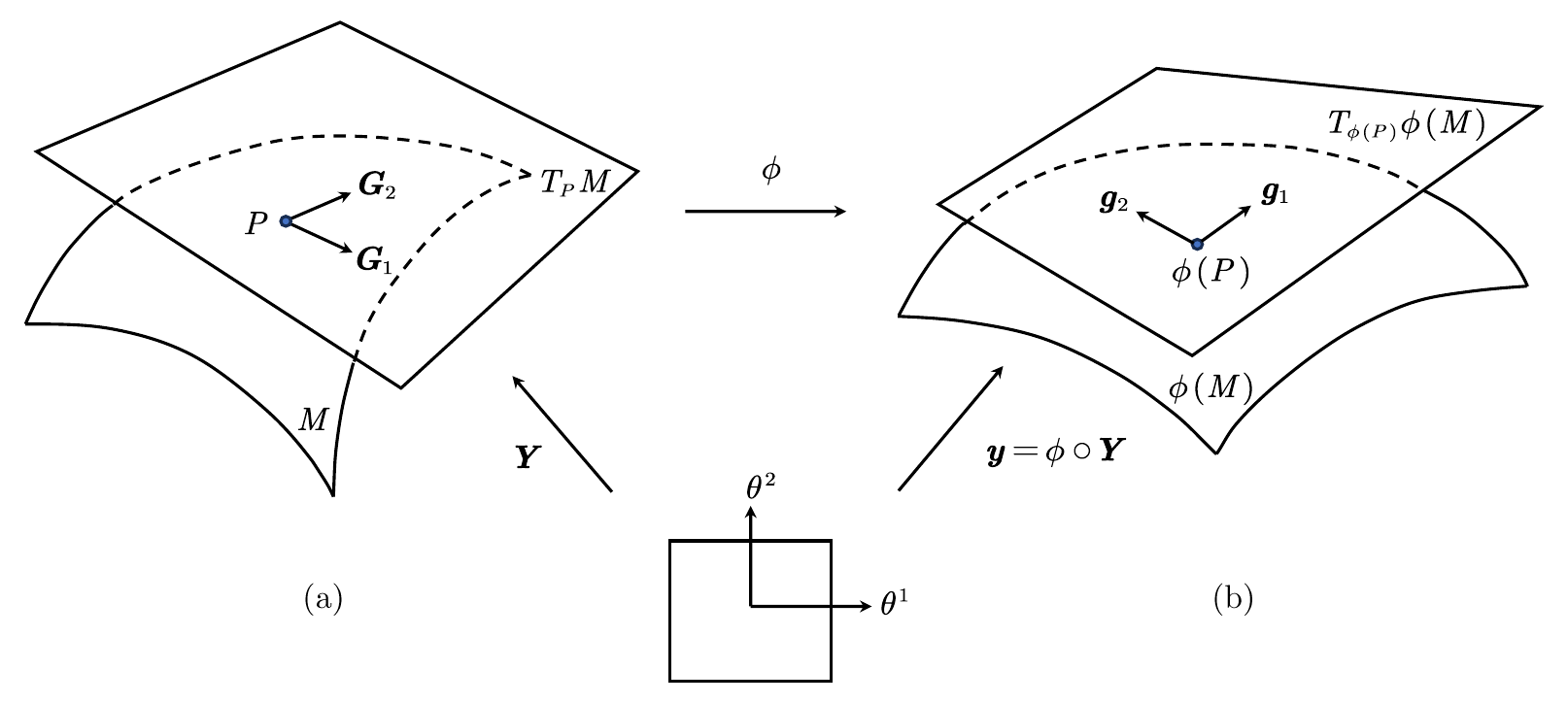}
	\caption{An elastic surface in (a) reference configuration (before deformation) and (b) current configuration (after deformation).}
	\label{fig:surface}
\end{figure}

We assume that the bulk solid is subjected to surface stresses on a smooth part $M$ of the boundary $\partial \Omega$, and prescribed (dead-load) traction $\hat{\bm{t}}$ and position $\hat{\bm{a}}$ are imposed on the remaining parts $\partial\Omega_t$ and $\partial \Omega_u$ of the boundary, respectively. The stressed surface $M$ is parameterized locally by  curvilinear coordinates $\theta^1$ and $\theta^2$. The position function $\bm{Y}$ on the surface  is identified with the restriction of $\bm{X}$ to $M$,
\begin{align}\label{eq:M}
\bm{Y}(\theta^1,\theta^2)=\bm{X}|_{M}.
\end{align}
The surface $M$ is assumed to be convected by the deformation of the solid so that its image  $\phi(M)$ under the deformation admits the local parametrization
\begin{align}\label{eq:chiM}
\bm{y}(\theta^1,\theta^2)=\phi(\bm{Y}(\theta^1,\theta^2)),
\end{align}
as shown in Fig. \ref{fig:surface}.
This parametrization induces the tangent vectors
\begin{align}\label{eq:Gg}
\bm{G}_{\alpha}=\bm{Y}_{,\alpha}\in T_PM,\quad \bm{g}_{\alpha}=\bm{y}_{,\alpha}\in T_{\phi(P)}\phi(M),
\end{align}
where $T_PM$ and $T_{\phi(P)}\phi(M)$ are the tangent planes to the surfaces $M$ and $\phi(M)$ at the points $P=\bm{Y}(\theta^1,\theta^2)$ and $\phi(P)$, respectively, and as stated earlier a comma indicates partial differentiation with respect to the corresponding curvilinear coordinate.

The oriented unit normals to the tangent planes $T_PM$ and $T_{\phi(P)}\phi(M)$  are defined by
\begin{align}
\bm{N}=\frac{\bm{G}_1\times \bm{G}_2}{\sqrt{G}},\quad \bm{n}=\frac{\bm{g}_1\times \bm{g}_2}{\sqrt{g}},
\end{align}
where $G=\det(\bm{G}_\alpha\cdot\bm{G}_\beta)$ and $g=\det(\bm{g}_\alpha\cdot\bm{g}_\beta)$ are the metric determinants  with $(\bm{g}_\alpha\cdot\bm{g}_\beta)$, for instance,  denoting the $2\times 2$ matrix whose $\alpha\beta$-component is the dot product $\bm{g}_\alpha\cdot\bm{g}_\beta$. With the aid of the unit normals, the dual tangent vectors induced by $\theta^\alpha$ can be expressed as
\begin{align}
\bm{G}^\alpha=\frac{e^{\alpha\beta}}{\sqrt{G}}\bm{G}_\beta\times \bm{n},\quad \bm{g}^\alpha=\frac{e^{\alpha\beta}}{\sqrt{g}}\bm{g}_\beta\times \bm{n},
\end{align} where $e^{\alpha\beta}=e_{\alpha\beta}$ is the unit alternator ($e^{12}=-e^{21}=1$, $e^{11}=e^{22}=0)$. These dual vectors satisfy the relations $\bm{G}^\alpha\cdot\bm{G}_\beta=\bm{g}^\alpha\cdot\bm{g}_\beta=\delta^\alpha_\beta$, where $\delta^\alpha_\beta$ is the Kronecker delta. The variations of the unit normals are described by the curvature tensors:
\begin{align}\label{eq:Kk}
\bm{B}=-\frac{\partial\bm{N}}{\partial\bm{Y}}=-\bm{N}_{,\alpha}\otimes \bm{G}^\alpha,\quad \bm{b}=-\frac{\partial\bm{n}}{\partial\bm{y}}=-\bm{n}_{,\alpha}\otimes \bm{g}^\alpha.
\end{align}

According to the chain rule, the local basis vectors in the reference and current configurations are related by
\begin{align}
\bm{g}_\alpha=\bm{F}\bm{G}_{\alpha}.
\end{align}
However, it is useful to have a description of the deformation of the surface that does not involve the deformation of the bulk solid. This can be achieved by introducing the {\it surface deformation gradient}
\begin{align}\label{eq:Fs}
\bm{F}_s=\frac{\partial\bm{y}}{\partial\bm{Y}}=\bm{y}_{,\alpha}\otimes \bm{G}^\alpha=\bm{g}_{\alpha}\otimes \bm{G}^\alpha=\nabla\bm{y},
\end{align}
where here and henceforth the subscript `s' refers to quantities associated with the surface,  and the last equation in \rr{eq:Fs} serves to define the surface del operator $\nabla=\bm{G}^\alpha\frac{\partial}{\partial\theta^\alpha}$ on the undeformed surface $M$.
On the other hand, the deformation gradient of the bulk material can be expressed as
\begin{align}\label{eq:Fbulk}
\bm{F}=\bm{x}_{,\alpha}\otimes \bm{G}^\alpha+\bm{x}_{,N}\otimes \bm{N},
\end{align}
where $\bm{x}_{,N}:=\frac{d}{dt}\bm{x}(\bm{X}+t\bm{N})|_{t=0}$ denotes the directional derivative of $\bm{x}$ along the direction given by $\bm{N}$. Noting that  $\bm{y}=\bm{x}|_{\phi(M)}$ is the restriction of $\bm{x}$ on the deformed surface $\phi(M)$, we deduce from \eqref{eq:Fs} and \eqref{eq:Fbulk} that
\begin{align}\label{eq:A}
\bm{F}_s=\bm{F}\bm{1} \;\;\;\; {\rm on}\;\; M,
\end{align}
where $\bm{1}=\bm{I}-\bm{N}\otimes \bm{N}$ denotes the projection tensor to the tangent plane of $M$, which is also the identity tensor on the same	 plane and $\bm{I}$ is the identity tensor on $\mathbb{R}^3$. A direct consequence of  \eqref{eq:A} is that the surface deformation gradient is {\it superficial} in the sense that it possesses the property $\bm{F}_s\bm{N}=0$ \citep{javili2018aspects}. The superficial property plays an important role in the surface elasticity theory.

It follows from \rr{eq:Fs} that $\bm{n}\cdot \bm{F}_s=0$. Thus, $\bm{F}_s$ is essentially a two-point plane tensor in the sense that it can be identified with a $2\times 2$ matrix when referred to bases of the tangent planes $T_PM$ and $T_{\phi(P)}\phi(M)$. Therefore, the determinant of ${\bm F}_s$ may be defined by
\begin{align}
&|\bm{F}_s{\bm u} \times \bm{F}_s {\bm v}|=\det(\bm{F}_s) |{\bm u} \times {\bm v}|, \quad \forall\,\bm{u}, \bm{v}\in T_PM. \la{added2}
\end{align}
This generalizes the definition of determinant for tensors defined on the plane $ \mathbb{R}^2$ in the language of exterior algebra \citep{winitzki2009linear}, and is consistent with the formula for the determinant of $2\times 2$ matrices when $\bm{F}_s$ is referred to orthonormal bases of the tangent planes $T_PM$ and $T_{\phi(P)}\phi(M)$.

Suppose that $\bm{F}_s$ depends on a parameter $t$. With the use of the same method given by \cite{chadwickbook} for tensors defined on $ \mathbb{R}^3$, it can be shown that
Jacobi's formula
\begin{align}\la{added3}
\frac{d J_s}{d t} =J_s \tr \Big ( \frac{d \bm{F}_s}{d t} \bm{F}_s^{-1}\Big)
\end{align}
is still valid, where $J_s=\det(\bm{F}_s)$ and $\bm{F}_s^{-1}$ is the inverse of $\bm{F}_s$ with the latter viewed as a linear transformation from the two-dimensional vector space $T_PM$ to its deformed counterpart $T_{\phi(P)}\phi(M)$. In particular, we have
\begin{align}\label{eq:Js}
\frac{\partial J_s}{\partial \bm{F}_s}= J_s\bm{F}_s^{-T},
\end{align}
where $\bm{F}_s^{-T}$ denotes the transpose of $\bm{F}_s^{-1}$. Note that by definition, we have
\begin{align}\label{july1}
\bm{F}_s^{-1}\bm{F}_s=\bm{1},\quad \bm{F}_s\bm{F}_s^{-1}=\bm{i},
\end{align}
where $\bm{i}=\bm{I}-\bm{n}\otimes \bm{n}$ stands for the  spatial identity tensor.

\section{Finite-strain surface elasticity theory}\label{sec:set}
In this section, we summarize and adapt slightly the surface elasticity theory given by Steinmann \cite{steinmann2008boundary}.
\subsection{Constitutive  relations of elastic surfaces}
The surface elasticity theory assumes that a surface has its own constitutive relation which is derived from the variation of the surface energy. Let us denote by $\varGamma$ the surface energy per unit reference area. We assume that  $\varGamma$ is a function of the surface deformation gradient, i.e., that the surface is hyperelastic.   Then the surface energy of the solid is given by
\begin{align}\label{eq:es}
\mathcal{E}_s=\int_{M} \varGamma(\bm{F}_s)\,dA,
\end{align}
Applying variations to \eqref{eq:es}, we obtain
\begin{align}\label{eq:variation}
\delta\mathcal{E}_s=\int_{M} \bm{P}_s:\delta \bm{F}_s\,dA,
\end{align}
where $\bm{P}_s$ is first Piola-Kirchhoff (P--K) stress  of the surface and is defined by
\begin{align}\label{eq:Ss}
\bm{P}_s:=\frac{\partial \varGamma}{\partial \bm{F}_s}.
\end{align}

Eq. \eqref{eq:Ss} gives the constitutive relation of the surface. In particular, for an isotropic surface, $\varGamma$ is only a function of two invariants  $I^s_1=\tr(\bm{C}_s)$ and $I^s_2=\det(\bm{C}_s)=J_s^2$ of the right Cauchy--Green tensor $\bm{C}_s=\bm{F}_s^T\bm{F}_s$ of the surface.  Then using  \eqref{eq:Js}, the first P--K stress of the surface is obtained as
\begin{align}\label{eq:Ps}
\bm{P}_s=2\frac{\partial \varGamma}{\partial I^s_1}\bm{F}_s +J_s \frac{\partial \varGamma}{\partial J_s}\bm{F}_s^{-T}.
\end{align}
In the Eulerian description, the Cauchy stress of the surface can be expressed as
\begin{align}\label{eq:sigmas}
\bm{\sigma}_s=J_s^{-1}\bm{P}_s\bm{F}_s^{T}=2J_s^{-1}\frac{\partial \varGamma}{\partial I_1^s}\bm{F}_s\bm{F}_s^T+\frac{\partial \varGamma}{\partial J_s}\bm{i}.
\end{align}
The linearization of \eqref{eq:sigmas} yields the well-known Shuttleworth equation \citep{shuttleworth1950surface}.

\subsection{Equilibrium equations of elastic surfaces}

Assume that the bulk solid is composed of a hyperelastic material, with the  strain energy function $W(\bm{F})$. The total potential energy of the solid is the sum of the bulk energy,  surface energy and load potential:
\begin{align}\label{eq:totale}
\mathcal{E}=\int_\Omega W(\bm{F})\,dV+\int_{M} \varGamma(\bm{F}_s)\,dA-\int_{\partial\Omega_t}\hat{\bm{t}}\cdot\bm{x}\,dA,
\end{align}
where $\partial\Omega_t$ is a part of $\partial\Omega$ on which traction $\hat{\bm{t}}$ is prescribed.

According to the principle of stationary potential energy, the equilibrium equations of the elastic solid are obtained by setting the first variation of the total potential energy to zero. By the divergence theorem,  the variation of the bulk energy $\mathcal{E}_b=\int_\Omega W(\bm{F})\,dV$ is calculated as
\begin{align}\label{eq:deltaeb}
\delta\mathcal{E}_b=\int_{\partial \Omega}\bm{P}\bm{N}\cdot\delta\bm{x}\,dA-\int_{\Omega}\Div(\bm{P}^T)\cdot\delta\bm{x}\,dV,
\end{align}
where $\bm{P}=\frac{\partial W}{\partial\bm{F}}$ is the first P--K stress of the bulk and $\Div$ denotes the divergence with respect to the Lagrangian coordinates $\bm{X}=X_j\bm{e}_j$ (i.e.,  $\Div(\bm{T}):=T_{ij,i}\bm{e}_j$).

Substituting \eqref{eq:Fs} into \eqref{eq:variation} and noting that $\bm{y}=\bm{x}|_{\phi(M)}$, we see that the variation of the surface energy $\mathcal{E}_s$ is
\begin{align}\label{eq:deltaEs}
\begin{split}
\delta\mathcal{E}_s&=\int_{M}\bm{P}_s: (\delta\bm{x}_{,\alpha}\otimes\bm{G}^\alpha)\,dA=\int_{M} \bm{G}^\alpha\cdot \bm{P}_s^T\delta\bm{x}_{,\alpha}\,dA\\
&=\int_{M} (\bm{G}^\alpha\cdot (\bm{P}^T_s\delta\bm{x})_{,\alpha}-\bm{G}^\alpha\cdot \bm{P}^T_{s,\alpha}\delta\bm{x})\,dA\\
&=\int_{M} (\nabla\cdot (\bm{P}_s^T\delta\bm{x})-(\nabla\cdot\bm{P}^T_s)\cdot\delta\bm{x})\,dA,
\end{split}
\end{align}
where $\nabla\cdot(\bullet)=\bm{G}^\alpha \cdot (\bullet)_{,\alpha}$ signifies the surface divergence.
Since $\bm{P}_s$ is conjugate to $\bm{F}_s$ and $\bm{F}_s$ is superficial, it follows that $\bm{P}_s$ is also superficial, i.e., $\bm{P}_s\bm{N}=0$. This implies that $\bm{P}_s^T\delta\bm{x}$ is a vector that lies on the tangent plane of $M$, so we can apply the surface divergence theorem (\cite{steinmann2008boundary}, Eq. (12)) to obtain
\begin{align}\label{eq:deltaes}
\delta\mathcal{E}_s=\int_{\partial M} \bm{P}_s\bm{V}\cdot \delta\bm{x}\,dS-\int_{M}(\nabla\cdot\bm{P}_s^T)\cdot\delta\bm{x}\,dA,
\end{align}
where $\partial M$ denotes the boundary of $M$ and $\bm{V}$ is its unit outward normal. 

Adding \eqref{eq:deltaeb} and \eqref{eq:deltaes} together, we see that the variation of the total potential energy is given by
\begin{align}
\begin{split}
\delta\mathcal{E}=&-\int_{\Omega}\Div(\bm{P}^T)\cdot\delta\bm{x}\,dV+\int_{ \partial\Omega_t}(\bm{P}\bm{N}-\hat{\bm{t}})\cdot\delta\bm{x}\,dA+\int_{ \partial\Omega_u}\bm{P}\bm{N}\cdot\delta\bm{x}\,dA\\
&+\int_{M}(\bm{P}\bm{N}-\nabla\cdot\bm{P}_s^T)\cdot\delta\bm{x}\,dA+\int_{\partial M}\bm{P}_s\bm{V}\cdot\delta\bm{x}\,dS.
\end{split}
\end{align}
Setting $\delta \mathcal{E}=0$ then yields the following equilibrium equations and associated boundary conditions of the surface elasticity theory:
\begin{align}
&\Div(\bm{P}^T)=0\quad \text{in}\ \Omega, \label{eq:e1}\\
&\bm{P}\bm{N}=\hat{\bm{t}}\quad \text{on}\ \partial  \Omega_t, \label{eq:e2}\\
&\bm{x}=\hat{\bm{a}}\quad \text{on}\ \partial\Omega_u, \label{eq:e2add}\\
&\bm{P}\bm{N}=\nabla\cdot\bm{P}^T_s\quad \text{on}\ M, \label{eq:e3}\\
&\bm{P}_s\bm{V}=0 \quad \text{on}\ \partial M. \label{eq:e4}
\end{align}
In particular, Eq. \eqref{eq:e3} describes the equilibrium between the surface stress and the stress in the bulk, which is usually called the  generalized Young--Laplace equation. One can check that \eqref{eq:e3} agrees with Eq. (4.21) in \cite{steigmann1999elastic} and Eq. (59) in \cite{gao2014curvature} when the surface bending stiffness is neglected. Note that Steigmann and Ogden \cite{steigmann1999elastic} used a different definition of the surface divergence operator, which can be proved to be equivalent to ours by employing the fact that the divergence of a field is equal to  the trace of  its gradient.

Using the same variational technique, one can derive from \eqref{eq:totale} the following Eulerian form of \eqref{eq:e3}:
\begin{align}\label{eq:YL}
\bm{\sigma}\bm{n}=\tilde{\nabla}\cdot\bm{\sigma}_s\quad \text{on}\ \phi(M),
\end{align}
where $\bm{\sigma}$ denotes the Cauchy stress of the bulk and $\tilde{\nabla}\cdot\bm{\sigma}_s = \bm{g}^\alpha\cdot \bm{\sigma}_{s,\alpha}$ denotes the spatial divergence of
$\bm{\sigma}_s$.
In the special case where $\varGamma= \gamma J_s$ with $\gamma$ being a constant (i.e., liquid-like surface tension),  Eqs. \rr{eq:sigmas} and \eqref{eq:Kk} may be used to evaluate $\tilde{\nabla}\cdot\bm{\sigma}_s$. After some simplification, Eq.
\rr{eq:YL} reduces to
\begin{align}
\bm{\sigma}\bm{n}=\gamma\tr(\bm{b})\bm{n} \quad \text{on}\ \phi(M),
\la{kappa}
\end{align}
which is the familiar Young--Laplace equation \citep{taffetani2015beading}.

\section{Incremental equations in surface elasticity theory and their linearization}\label{sec:incremental}

This section derives the exact equations governing incremental deformations  superimposed on a finite deformation in surface elasticity theory and the  corresponding linearized forms.

\subsection{Exact incremental equations}
We follow Ogden \cite{ogden1984non} who established a general theory of small deformations superimposed on a finite deformation in an elastic material. Let us denote by $B_0$ the initially unstressed configuration of the elastic solid which occupies the region $\Omega\subset\mathbb{R}^3$.  A smooth part $M$ of the boundary $\partial \Omega$ of the solid is subjected to surface stresses, and the rest of the boundary is subjected to either prescribed traction or displacement, giving rise to a finitely stressed equilibrium configuration $B_e$. Finally, a displacement (not necessarily small) is superimposed on $B_e$, resulting in a configuration, termed the current configuration and denoted by $B_t$. The position vectors of a material point in $B_0$, $B_e$ and $B_t$ are denoted by $\bm{X}$, $\bm{x}$ and $\tilde{\bm{x}}$, respectively. Then we have
\begin{align}
\tilde{\bm{x}}(\bm{X})=\bm{x}(\bm{X})+\bm{u}(\bm{x}),
\end{align}
where $\bm{u}$ is the displacement superimposed on $B_e$. For easy reference, we use $\phi$ to denote the deformation map from $B_0$ to $B_e$, so that  $\bm{x}=\phi(\bm{X})$ and $\phi(M)$ represents the finitely deformed surface in  $B_e$.

The deformation gradients arising from the deformations $B_0\to B_t$ and $B_0\to B_e$ are denoted by $\bm{F}$ and $\bar{\bm{F}}$ respectively, which are given by
\begin{align}\label{eq:FFbar}
\bm{F}=\frac{\partial\tilde{\bm{x}}}{\partial\bm{X}},\quad \bar{\bm{F}}=\frac{\partial\bm{x}}{\partial\bm{X}}.
\end{align}
It follows from the chain rule that
\begin{align}\label{eq:FF}
\bm{F}=(\bm{I}+\bm{\eta})\bar{\bm{F}},
\end{align}
where $\bm{\eta}=\frac{\partial\bm{u}}{\partial\bm{x}}$ is the incremental displacement gradient. From \eqref{eq:A}, the  surface counterpart of \eqref{eq:FF} is given by
\begin{align}\label{eq:Fetas}
\bm{F}_s=(\bar{\bm{i}}+\bm{\eta}_s)\bar{\bm{F}}_s,
\end{align}
where  $\bar{\bm{F}}_s=\bar{\bm{F}} {\bm 1}$, $\bar{\bm{i}}$ denotes the identity tensor on the tangent plane of $\phi(M)$, and $\bm{\eta}_s=\bm{\eta}\bar{\bm{i}}$ is the incremental surface displacement gradient.

Let $\bm{N}\,dA$ denote a vector surface area element in $B_0$, where $\bm{N}$ is  the unit outward normal to the surface, and $\bm{n}\,da$ the corresponding area element in $B_e$. The area elements are connected by Nanson's formula
\begin{align}\label{eq:nanson}
\bm{N}\,dA=\bar{J}^{-1}\bar{\bm{F}}^T\bm{n}\,da,
\end{align}
where $\bar{J}=\det(\bar{\bm{F}})$. Similarly, let $\bm{V}\,dS$ denote a vector line element in $B_0$, where $\bm{V}$ is the unit outward normal to the line, and $\bm{v}\,ds$ the corresponding line element in $B_e$. The line elements are connected by the surface Nanson formula (\cite{steinmann2008boundary}, Eq. (49))
\begin{align}\label{eq:2dnanson}
\bm{V}\,dS=\bar{J}_s^{-1}\bm{\bar{F}}_s^{T}\bm{v}\,ds,
\end{align}
where $\bar{J}_s=\det(\bar{\bm{F}}_s)$.

The first P--K stresses associated with the deformations $B_0\to B_t$ and $B_0\to B_e$ are denoted by $\bm{P}$ and $\bar{\bm{P}}$, and are given by
\begin{align}\label{eq:SSbar}
\bm{P}=\frac{\partial W}{\partial\bm{F}},\quad \bar{\bm{P}}=\frac{\partial W}{\partial \bm{F}}\Big|_{\bar{\bm{F}}}.
\end{align}
Their surface counterparts  take the form
\begin{align}\label{eq:SSsbar}
\bm{P}_s=\frac{\partial\varGamma}{\partial\bm{F}_s},\quad \bar{\bm{P}}_s=\frac{\partial \varGamma}{\partial\bm{F}_s}\Big|_{\bar{\bm{F}}_s},
\end{align}
where $\bm{P}_s$ and $\bar{\bm{P}}_s$ are the surface first  P--K stresses corresponding to the deformations $B_0\to B_t$ and $B_0\to B_e$, respectively.

To state the incremental equations, it is convenient to introduce an incremental stress tensor  and its surface counterpart through
\begin{align}\label{eq:chi}
&\bm{\chi}=\bar{J}^{-1}(\bm{P}-\bar{\bm{P}})\bar{\bm{F}}^T, \quad\bm{\chi}_s=\bar{J}_s^{-1}(\bm{P}_s-\bar{\bm{P}}_s)\bar{\bm{F}}_s^T.
\end{align}
The incremental forms of \eqref{eq:e1}--\eqref{eq:e2add} have been given in \cite{ogden1984non} and are of the form
\begin{align}
&\ddiv(\bm{\chi}^T)=0\quad \text{in}\ \phi(\Omega), \\
&\bm{\chi}\bm{n}=0\quad \text{on}\  \phi(\partial\Omega_t),\\
&\bm{u}=0\quad \text{on}\  \phi(\partial\Omega_u),
\end{align}
where $\ddiv$ means the divergence with respect to coordinates $\bm{x}=x_j\bm{e}_j$ in $B_e$ (i.e., $\ddiv(\bm{T}):=T_{ij,i}\bm{e}_j$) 

To derive the incremental equations of \eqref{eq:e3} and \eqref{eq:e4},  we apply increments to  \eqref{eq:e3} to obtain
\begin{align}\label{eq:i3}
(\bm{P}-\bar{\bm{P}})\bm{N}=\nabla\cdot (\bm{P}_s-\bar{\bm{P}}_s)^T\quad \text{on}\ M.
\end{align}
To obtain the parallel equation of \eqref{eq:i3} valid on the finitely deformed surface $\phi(M)$,  let $R$ an arbitrary region of $M$ and integrating \eqref{eq:i3} over $R$ yields
\begin{align}
\int_{R} (\bm{P}-\bar{\bm{P}})\bm{N}\,dA=\int_{R} \nabla\cdot(\bm{P}_s-\bar{\bm{P}}_s)^T\,dA.
\end{align}
Since $\bm{P}_s-\bar{\bm{P}}_s$ is superficial, by Nanson's formula and the surface divergence theorem, we have
\begin{align}\label{eq:chiR}
\int_{\phi(R)} \bar{J}^{-1}(\bm{P}-\bar{\bm{P}})\bar{\bm{F}}^T\bm{n}\,da=\int_{\partial R}(\bm{P}_s-\bar{\bm{P}}_s)\bm{V}\,dS.
\end{align}
Applying the surface Nanson formula \eqref{eq:2dnanson}, we can rewrite \eqref{eq:chiR} as
\begin{align}
\int_{\phi(R)} \bm{\chi}\bm{n}\,da=\int_{\partial\phi(R)}\bar{J}_s^{-1}(\bm{P}_s-\bar{\bm{P}}_s)\bar{\bm{F}}_s^T\bm{v}\,ds=\int_{\partial\phi(R)}\bm{\chi}_s\bm{v}\,ds.
\end{align}
Then another application of the surface divergence theorem yields
\begin{align}
\int_{\phi(R)} \bm{\chi}\bm{n}\,da=\int_{\phi(R)}\varnabla\cdot\bm{\chi}_s^T\,da,
\end{align}
where $\varnabla=\bm{g}^\alpha\frac{\partial}{\partial\theta^\alpha}$ is the surface del operator on $\phi(M)$ and $\bm{g}^\alpha$ are the tangent vectors of $\phi(M)$ induced by the curvilinear coordinates $\theta^\alpha$. Since $R$ is an arbitrary region of $M$, we conclude that
\begin{align}
\bm{\chi}\bm{n}=\varnabla\cdot\bm{\chi}_s^T \quad \text{on}\  \phi(M).
\end{align}
A similar argument shows that the incremental equation of \eqref{eq:e4} is
\begin{align}
\bm{\chi}_s\bm{v}=0 \quad \text{on}\  \phi(\partial M).
\end{align}

On collecting all the incremental equations together, we have
\begin{align}
&\ddiv(\bm{\chi}^T)=0\quad \text{in}\ \phi(\Omega), \label{eq:chi1}\\
&\bm{\chi}\bm{n}=0\quad \text{on}\ \phi(\partial\Omega_t), \label{eq:chi2}\\
& \bm{u}=0 \quad \text{on}\  \phi(\partial\Omega_u),\\
&\bm{\chi}\bm{n}=\varnabla\cdot\bm{\chi}_s^T \quad \text{on}\  \phi(M), \label{eq:chi3}\\
&\bm{\chi}_s\bm{v}=0 \quad \text{on}\  \phi(\partial M). \label{eq:chi4}
\end{align}
We remark that \eqref{eq:chi2} can be easily adapted to some non-dead loading cases such as pressure loading \citep{haughton1979bifurcation}.

\subsection{Linearized incremental equations}

To state the results, we choose  an orthonormal basis   $\{\bm{e}_1,\bm{e}_2,\bm{e}_3\}$ for $\mathbb{R}^3$ such that $\{\bm{e}_1,\bm{e}_2\}$ span the tangent plane of finitely deformed surface $\phi(M)$ and $\bm{e}_3=\bm{n}$, the unit normal to $\phi(M)$.  The components of a tensor $\bm{T}$ relative to the  basis $\{\bm{e}_1,\bm{e}_2,\bm{e}_3\}$  is denoted by $T_{ij}=\bm{e}_i\cdot \bm{T}\bm{e}_j$, and all the results presented in the following will be expressed in terms of the components relative to this chosen basis.

Assume that the displacement $\bm{u}$ is small so that terms of order $|\bm{u}|^2$ can be neglected. Substituting \eqref{eq:SSbar} into $\eqref{eq:chi}_1$ and linearizing around $\bm{F}=\bar{\bm{F}}$ with the use of \eqref{eq:FF}, one obtains the following linearized stress-strain relation for incremental deformation of the bulk
\begin{align}\label{eq:chiij}
\chi_{ij}=\mathcal{A}_{jilk}\eta_{kl},
\end{align}
where  $\A_{jilk}$  are the first-order instantaneous moduli  given by
\begin{align}\label{eq:calB}
\mathcal{A}_{jilk}=\bar{J}^{-1}\bar{F}_{jm}\bar{F}_{ln}\frac{\partial^2 W}{\partial F_{im}\partial F_{kn}}\Big|_{\bar{\bm{F}}}.
\end{align}
In a similar spirit, substituting \eqref{eq:SSsbar} into $\eqref{eq:chi}_2$ and linearizing around $\bm{F}_s=\bar{\bm{F}}_s$  with the use of \eqref{eq:Fetas}, we obtain the following linearized stress-strain relation for incremental deformations of the surface
\begin{align}\label{eq:chis}
\chi^s_{{i\alpha}}=\mathcal{C}_{\alpha i \beta j} \eta^s_{j\beta},
\end{align}
where we have used the fact that $\chi^s_{i3}=\eta^s_{i3}=0$ which follows from  the superficial property of $\bm{\chi}_s$ and $\bm{\eta}_s$ with respect to $\bm{n}$,  and $\mathcal{C}_{\alpha i\beta j}$ are the first-order instantaneous moduli of the surface defined by
\begin{align}\label{eq:calC}
\mathcal{C}_{\alpha i \beta j}=\bar{J}^{-1}_s \bar{F}^s_{\alpha k}\bar{F}^s_{\beta l}\frac{\partial^2 \varGamma}{\partial F^s_{ik }\partial F^s_{jl}}\Big|_{\bar{\bm{F}}_s}.
\end{align}
Note that in the above expressions we have written $\bm{\chi}^s$, $\bm{\eta}^s$ and $\bar{\bm{F}}^s$ for $\bm{\chi}_s$, $\bm{\eta}_s$ and $\bar{\bm{F}}_s$, respectively, when their components are displayed for better readability.  This convention will be adopted in  later sections. We remark that the components $\bar{F}^s_{3j}$ do not appear  in \eqref{eq:calC} because they are  zero, a consequence of the equality $\bm{n}\cdot\bar{\bm{F}}_s=0$.  In contrast, the components $F^s_{i3}$ are not zero since $\bm{F}_s$ is not superficial with reference to $\bm{n}$. The linearized governing equations are obtained by substituting the linearized stress-strain relations \eqref{eq:calB} and \eqref{eq:calC} into the  governing equations \eqref{eq:chi1}--\eqref{eq:chi4}.

\subsection{Surface moduli for isotropic materials}\label{sec:surface}
When the bulk solid and the surface are both isotropic in their undeformed configuration, more explicit expressions can be obtained in terms of the principal stretches.
We may assume that for the deformation from $B_0$ to $B_e$,  the basis vectors $\{\bm{e}_1, \bm{e}_2, \bm{e}_3\}$ employed in the previous section coincide with the principal axes of stretch in the bulk solid, and $\{\bm{e}_1, \bm{e}_2\}$ coincide with the principal axes of stretch on the surface. Let $\lambda_1$, $\lambda_2$ and $\lambda_3$ denote the principal stretches of the deformation from $B_0$ to $B_e$, corresponding to the principal directions $\bm{e}_1$, $\bm{e}_2$ and $\bm{e}_3$, respectively. For the bulk solid, the expressions for the nonzero components of the first-order instantaneous modulus $\mathcal{A}$ have been given in \rr{eqn3}.

In the isotropic case, the surface energy $\varGamma=\varGamma(\lambda_1^s,\lambda_2^s)$ is a function of the two principal surface stretches $\lambda_1^s$ and $\lambda_2^s$. Using the chain rule, we deduce from \eqref{eq:calC} that  for isotropic surfaces the nonzero components of the first-order surface instantaneous moduli $\mathcal{C}$  are simply (see \hyperref[app:A]{Appendix A} for details)
\begin{align}
\begin{split}\label{eq:C1}
&\bar{J}_s \mathcal{C}_{\alpha\alpha\beta\beta}=\lambda_{\alpha}\lambda_\beta\varGamma_{\alpha\beta},  \\
&\bar{J}_s\mathcal{C}_{\alpha\beta\alpha\beta}=\lambda_\alpha^2\frac{\lambda_\alpha \varGamma_\alpha-\lambda_\beta\varGamma_\beta}{\lambda_\alpha^2-\lambda_\beta^2},
\quad \alpha\ne \beta, \\
&\bar{J}_s\mathcal{C}_{\alpha\beta\beta\alpha}=\lambda_\alpha\lambda_\beta\frac{\lambda_\beta \varGamma_\alpha-\lambda_\alpha\varGamma_\beta}{\lambda_\alpha^2-\lambda_\beta^2}=\bar{J}_s\mathcal{C}_{\alpha\beta\alpha\beta}-\lambda_\alpha\varGamma_\alpha,\quad \alpha\ne \beta, \\
&\bar{J}_s\mathcal{C}_{\alpha3\alpha3}=\lambda_\alpha\varGamma_\alpha, 
\end{split}
\end{align}
where $\bar{J}_s=\lambda_1\lambda_2$,  $\varGamma_\alpha=\frac{\partial\varGamma}{\partial \lambda_\alpha^s}|_{(\lambda_1,\lambda_2)}$ and $\varGamma_{\alpha\beta}=\frac{\partial^2\varGamma}{\partial \lambda_\alpha^s\partial\lambda^s_\beta}|_{(\lambda_1,\lambda_2)}$. The last equality in $\eqref{eq:C1}_3$ reflects the fact that the dependence of $\varGamma$ on $\bm{F}_s$ is through $\bm{C}_s=\bm{F}_s^T\bm{F}_s$.

\section{Connection with the Gurtin--Murdoch model}\label{sec:5}

The well-known Gurtin--Murdoch model \citep{gurtin1975continuum,gurtin1978surface} expresses the surface stress tensor as a sum of an isotropic linear function of the surface strain and the residual stress induced by surface tension. Let $\gamma$, $\mu_s$ and $\lambda_s$ be the (residual) surface tension, shear modulus and Lame's first parameter of the surface, respectively. Then the surface stress tensor in the Gurtin--Murdoch model is given by (\cite{gurtin1978surface}, Eq. (2))
\begin{align}\label{eq:GM}
\bm{\sigma}^\text{GM}_s=\gamma\bm{1}+2(\mu_s-\gamma)\bm{\varepsilon}_s+(\lambda_s+\gamma)\tr(\bm{\varepsilon}_s)\bm{1}+\gamma \nabla\bm{u},
\end{align}
where $\bm{1}$ stands for the identity tensor on the tangent plane of the undeformed surface, $\bm{u}$ is the displacement, $\bm{\varepsilon}_s=\frac{1}{2}(\bm{1}\nabla\bm{u}+(\bm{1}\nabla\bm{u})^T)$ is the infinitesimal surface strain and $\nabla\bm{u}=\bm{u}_{,\alpha}\otimes \bm{G}^\alpha$ denotes the surface gradient of $\bm{u}$.

We now show that the Gurtin--Murdoch model can be recovered as a special case of the incremental theory derived in Section \ref{sec:incremental}. To this end, we consider the following surface energy function \citep{bakiler2023surface}
\begin{align}\label{eq:neos}
\varGamma=\gamma J_s +\frac{\mu_s}{2}(I^s_1-2-2\ln J_s)+\frac{\lambda_s}{2}\Big(\frac{1}{2}(J_s^2-1)-\ln J_s\Big),
\end{align}
which is a compressible neo-Hookean material model in surface elasticity. With the use of  \eqref{eq:chis} and \eqref{eq:calC},  it is direct to check that the incremental surface stress tensor $\bm{\chi}_s$ can be written in the compact form
\begin{align}\label{eq:chii}
\bm{\chi}_s=2(\mu_s^\text{eff}-\gamma)\bm{\varepsilon}_s+(\lambda_s^\text{eff}+\gamma)\tr(\bm{\varepsilon}_s)\bar{\bm{i}}+\bm{\eta}_s\bar{\bm{\sigma}}_s \quad (\text{present model}),
\end{align}
where
\begin{align}
& \mu_s^\text{eff}=\mu_s \bar{J}_s^{-1}-\frac{1}{2}\lambda_s(\bar{J}_s-\bar{J}_s^{-1}),\quad \lambda_s^\text{eff}=\lambda_s \bar{J}_s,\\
&\bm{\varepsilon}_s=\frac{1}{2}(\bar{\bm{i}}\bm{\eta}_s+(\bar{\bm{i}}\bm{\eta}_s)^T),\\
&\bar{\bm{\sigma}}_s=\gamma \bar{\bm{i}}+\mu_s \bar{J}_s^{-1}(\bar{\bm{F}}_s\bar{\bm{F}}_s^{T}-\bar{\bm{i}})+\frac{1}{2}\lambda_s(\bar{J}_s-\bar{J}_s^{-1})\bar{\bm{i}}.
\end{align}
The $\mu_s^\text{eff}$ and $\lambda_s^{\text{eff}}$ may be referred to as the effective surface shear modulus and Lame's first parameter, respectively, $\bm{\varepsilon}_s$ is obviously the incremental infinitesimal surface strain and $\bar{\bm{\sigma}}_s$ is the surface Cauchy stress associated with the deformation from $B_0$ to $B_e$.

The Gurtin--Murdoch model can be recovered from our incremental theory by assuming that the intermediate configuration $B_e$ coincides with the initial configuration $B_0$.  Under this assumption, we have $\bar{\bm{F}}_s=\bar{\bm{i}}=\bm{1}$, and consequently $\bm{\eta}_s=\nabla\bm{u}$ and $\bar{\bm{\sigma}}=\gamma\bm{1}$. Eq. \eqref{eq:chii} then reduces to
\begin{align}\label{eq:chiss}
\bm{\chi}_s=2(\mu_s-\gamma)\bm{\varepsilon}_s+(\lambda_s+\gamma)\tr(\bm{\varepsilon}_s)\bm{1}+\gamma \nabla\bm{u}.
\end{align}
By comparing \eqref{eq:GM} and \eqref{eq:chiss}, one can see that the total surface stress tensor of the present model, which is $\gamma\bm{1}+\bm{\chi}_s$, is the same as that of the Gurtin--Modorch model. In contrast with the Gurtin--Murdoch model which is only valid for small deformations, the current incremental theory (Eq. \eqref{eq:chii} in particular) is applicable even when the intermediate deformation is finite.

\section{Application to  an elastic cylinder subjected to surface stresses}\label{sec:6}
We now verify our incremental theory and demonstrate its ease of use by applying it to the bifurcation analysis of an elastic cylinder that is subjected to surface tension as well as a tensile and compressive  axial force.

When subjected to surface tension, an axially stretched soft cylinder may undergo a localized deformation culminating in a two-phase state, which is now commonly called the Plateau--Rayleigh instability \citep{taffetani2015beading} although in the context of fluid flows the same term refers to an instability mode with finite wavenumber. Previous stability analyses  \citep{taffetani2015beading, lestringant2020one, fu2021necking} have mainly focused on the case of constant surface tension, and have revealed  that the zero wavenumber mode is the dominant bifurcation mode. Very rarely has the situation when the surface tension is strain-dependent been examined, notable exceptions being the recent work of Bakiler \cite{bakiler2023surface}.  We now demonstrate that with the use of the incremental surface elasticity theory just established, we can study the latter problem more systematically. The axisymmetric buckling of a soft cylinder under compression is also analyzed as a by-product. This problem without surface tension has previously been examined by \cite{wilkes1955stability}. For convenience, we refer to
this instability in the presence of surface tension as the Wilkes instability, which has previously been studied by Wang \cite{wang2020axisymmetric}, and Emery and Fu \cite{emery2021localised} assuming that the surface tension is constant. Bifurcations of a coated elastic cylinder subjected to axial tension or compression have also been examined thoroughly by Bigoni and Gei \cite{bigoni2001bifurcations}.

\subsection{Homogeneous deformations}

We consider a hyperelastic solid cylinder with a radius $A$ in the reference configuration. The cylinder deforms homogeneously under the combined action of surface stresses on its outer surface and an axial force $N$ applied at its two ends. In terms of cylindrical coordinates, the homogeneous deformation is described by
\begin{align}\label{eq:hom}
r=aR,\quad \theta=\Theta,\quad z=\lambda Z,
\end{align}
where $(R,\Theta,Z)$ and $(r,\theta,z)$ are the cylindrical coordinates in the reference and current configurations, respectively, and $a$ and $\lambda$ are the constant transversal  stretch and axial stretch, respectively. 

From \eqref{eq:hom}, we see that the bulk deformation gradient is given by
\begin{align}\label{eq:Fb}
\bm{F}=a\bm{e}_{\theta}\otimes \bm{e}_{\theta}+\lambda\bm{e}_z\otimes \bm{e}_z+a\bm{e}_r\otimes \bm{e}_r,
\end{align}
where $(\bm{e}_r,\bm{e}_\theta,\bm{e}_z)$ is the common orthonormal basis for the two sets of cylindrical coordinates. The three principal stretches are simply
\begin{align}
\lambda_1=\lambda_3=a,\quad \lambda_2=\lambda,
\end{align}
where we have identified the indices $1,2,3$ with the $\theta$-, $z$- and $r$-directions, respectively. The surface deformation gradient can be obtained in a similar way, which is
\begin{align}
\bm{F}_s= \bm{F}\bm{1}=\bm{F}(\bm{I}-{\bm e}_r \otimes {\bm e}_r) =  a\bm{e}_{\theta}\otimes \bm{e}_{\theta}+\lambda\bm{e}_z\otimes \bm{e}_z.
\end{align}
The two principal stretches of the surface deformation are obviously given by
\begin{align}
\lambda_1^s=a=\lambda_1,\quad \lambda_2^s=\lambda=\lambda_2.
\end{align}

We assume that the constitutive law of the bulk solid is described by a strain energy function $W(\lambda_1,\lambda_2,\lambda_3)$ and that of the surface is characterized by a surface energy function
\begin{align}
\varGamma(\lambda_1^s,\lambda_2^s)=\gamma \lambda_1^s \lambda_2^s+\varPsi(\lambda_1^s,\lambda_2^s),
\end{align}
 where $\gamma$ signifies the surface tension and $\varPsi$ represents the strain-dependent part of the surface energy. Then the first P--K stress of the bulk is given by
\begin{align}
\bm{P}=W_1\bm{e}_\theta\otimes\bm{e}_\theta+W_2\bm{e}_z\otimes\bm{e}_z+W_3\bm{e}_r\otimes\bm{e}_r,
\end{align}
where $W_i=\partial W/\partial\lambda_i$.  Since the bulk stress is independent of the position, the equilibrium equations in the bulk  are satisfied automatically. Similarly, the surface first P--K stress  is given by
\begin{align}
\bm{P}_s=(\gamma\lambda^s_2+\varPsi_1)\bm{e}_{\theta}\otimes \bm{e}_{\theta}+(\gamma\lambda^s_1+\varPsi_2)\bm{e}_z\otimes \bm{e}_z,
\end{align}
where $\varPsi_\alpha=\partial\varPsi/\partial\lambda_\alpha^s$. The boundary condition \eqref{eq:e3} leads to
\begin{align}
P_{33}=-\frac{P^s_{11}}{A},
\end{align}
from which we obtain
\begin{align}\label{eq:gamma0a}
\gamma=-\frac{1}{2}A \lambda^{-1} w_1(a,\lambda)-\lambda^{-1}\varPsi_1(a,\lambda),
\end{align}
where $w(a,\lambda)=W(a,\lambda,a)$ is a reduced strain energy function and the subscript in $w$ indicates partial differentiation. Consequently, the resultant axial force at the two ends is given by
\begin{align}\label{eq:Nlambda}
N=\pi A^2 P_{22}+2\pi AP^s_{22}=\pi A^2 ( w_2(a,\lambda)-\lambda^{-1} a w_1(a,\lambda))+2 \pi A(\varPsi_2(a,\lambda)-  \lambda^{-1} a \varPsi_1(a,\lambda)).
\end{align}
With the aid of \rr{eq:gamma0a} and \rr{eq:Nlambda}, one can easily determine the deformation parameters $a$ and $\lambda$ once the loading parameters $\gamma$ and $N$ are given.

\subsection{Linear bifurcation analysis}

To study the bifurcation from the homogeneous deformation, we perturb the homogeneous deformation by adding a small displacement of the form
\begin{align}
\delta\bm{x}=u(r,z)\bm{e}_r+v(r,z)\bm{e}_z.
\end{align}
The incremental deformation gradient is calculated as
\begin{align}\label{eq:eta}
\bm{\eta}=\begin{pmatrix}
\frac{u}{r} & 0 & 0\\
0 & v_z & v_r\\
0 & u_z & u_r
\end{pmatrix},
\end{align}
where $v_z=\partial v/\partial z$, $v_r=\partial v/\partial r$, etc.

The identity tensor on the tangent space of the homogeneously deformed outer surface is   $\bar{\bm{i}}=\bm{e}_\theta\otimes \bm{e}_\theta+\bm{e}_z\otimes \bm{e}_z$. From the relation $\bm{\eta}_s=\bm{\eta}\bar{\bm{i}}$ on $r=aA$, we see that the surface deformation gradient $\bm{\eta}_s$ is given by
\begin{align}
\bm{\eta}_s=\begin{pmatrix}
\frac{u}{r} & 0 & 0\\
0 & v_z & 0\\
0 & u_z & 0
\end{pmatrix}\Bigg|_{r=aA}.
\end{align}
It is seen that the elements in the third column of $\bm{\eta}_s$ are all zero, which reflects the fact that $\bm{\eta}_s$  is superficial.

The linearized equilibrium equations \eqref{eq:chi1}  in cylindrical coordinates take the form
\begin{align}
&\frac{\partial\chi_{22}}{\partial z}+\frac{\partial\chi_{23}}{\partial r}+\frac{\chi_{23}}{r}=0, \label{eq:l2}\\
&\frac{\partial\chi_{33}}{\partial r}+\frac{\partial\chi_{32}}{\partial z}+\frac{\chi_{33}-\chi_{11}}{r}=0, \label{eq:l3}
\end{align}
where $(\chi_{ij})$ is the incremental stress tensor defined in \eqref{eq:chiij}.

According to \eqref{eq:chi3}, the incremental boundary conditions are given by
\begin{align}\label{eq:bc}
\chi_{23}=\frac{\partial\chi^s_{22}}{\partial z},\quad \chi_{33}=\frac{\partial\chi^s_{32}}{\partial z}-\frac{\chi^s_{11}}{r},\quad \text{on}\ r=aA,
\end{align}
where $(\chi^s_{_{ij}})$ is the incremental surface stress tensor that can be calculated using \eqref{eq:chis} and \eqref{eq:C1}. Information on the bifurcated solution in linear analysis is obtained by solving the boundary-value problem comprised of \eqref{eq:l2}--\eqref{eq:bc}.

For the numerical calculations carried out in the following,  we adopt the strain energy function
\begin{align}\label{eq:W}
W=\frac{\mu}{2}(I_1-3-2\ln J	)+\frac{\kappa}{2}\Big(\frac{1}{2}(J^2-1)-\ln J\Big)
\end{align}
for the bulk solid, where $\mu$ is the shear modulus, $\kappa=2\mu\nu/(1-2\nu)$ with $\nu\in [0,1/2]$ being Poisson's ratio, $I_1$ is the first principal invariant of the right Cauchy–Green tensor and $J$ is the determinant of the deformation gradient. For the surface, we assume that its energy function is given by \eqref{eq:neos}. We note that there are different choices of the last term in \eqref{eq:W} that are equivalent in the limit $\kappa\to\infty$. We pick the one that is in line with previous work \citep{dortdivanlioglu2022plateau,emery2023elasto,bakiler2023surface}. Also, for notational simplicity, we scale all length variables by $A$ and stress variables by  $\mu$, which is equivalent to set $A=1$ and $\mu=1$. We use the same letters to denote the scaled quantities. In particular, $\gamma$ and $N$ calculated using \eqref{eq:gamma0a} and \eqref{eq:Nlambda}  are now given by
\begin{align}
&\gamma=\frac{2+\kappa+2\mu_s+\lambda_s-2a^2-2a^2\mu_s-a^2\lambda^2\lambda_s-a^4\lambda^2\kappa}{2a\lambda}, \label{eq:gamma0sol}\\
&N=\frac{\pi(2+\kappa+2\lambda^2+4\lambda^2\mu_s-4a^2-4a^2\mu_s-a^4\lambda^2 \kappa)}{2\lambda}. \label{eq:Nsol}
\end{align}

To find the critical wavenumber of the bifurcation, we look for a normal mode solution of the form
\begin{align}\label{eq:uvp}
u(r,z)=f(r)e^{\ii kz},\quad v(r,z)=g(r)e^{\ii kz},
\end{align}
where $k$ is the axial wavenumber.  Substituting this solution into the incremental equations  \eqref{eq:l2}--\eqref{eq:bc} and simplifying the resulting equations with the use of \eqref{eqn3} and \eqref{eq:C1}, followed by eliminating $g(r)$ in favor of $f(r)$, we obtain a boundary value problem for $f(r)$:
\begin{align}
&\Big(\frac{d^2}{dr^2}+\frac{1}{r}\frac{d}{dr}-\Big(\frac{1}{r^2}+k^2q_1^2 \Big)\Big)\Big(\frac{d^2}{dr^2}+\frac{1}{r}\frac{d}{dr}-\Big(\frac{1}{r^2}+k^2 q_2^2\Big)\Big)f(r)=0,\label{eq:ode1}\\
& f'''(r)+b_1 f''(r)+b_2f'(r)+b_3 f(r)=0\quad \text{on} \ r=a,\label{eq:bc1}\\
& f'''(r)+\frac{2}{a}f''(r)+b_4f'(r)+b_5f(r)=0 \quad \text{on} \ r=a,\label{eq:bc2}
\end{align}
where
\begin{align}
&q_1=\frac{\lambda}{a},\quad q_2=\sqrt{\frac{2+\kappa+2\lambda^2+a^4\lambda^2\kappa}{2+\kappa+2a^2+a^4\lambda^2\kappa}},\\
&b_1=\frac{1}{a}\Big(2+\frac{2+\kappa+2\lambda^2+a^4\lambda^2 \kappa}{2\mu_s+\lambda_s+2\lambda^2\mu_s+a^2\lambda^2\lambda_s}\Big),
\end{align}
and the remaining coefficients $b_i$, $2\leq i\leq 5$ are available but are too long to be given here.

One can observe that the general solution to \eqref{eq:ode1} bounded at $r=0$  is given by
\begin{align}\label{eq:ff}
f(r)=c_1 I_1(kq_1r)+c_2 I_1(kq_2r),
\end{align}
where $I_1(x)$ denotes the modified Bessel function of the first kind, and $c_1$ and $c_2$ are arbitrary constants.
Substituting the above solution into \eqref{eq:bc1} and \eqref{eq:bc2}, we obtain a system of two linear equations in the unknowns $c_1$ and $c_2$. The existence of a nonzero solution requires that the determinant of the coefficient matrixmust vanish, which yields
\begin{align}\label{eq:dd}
\Omega(k,a,\lambda)=0,
\end{align}
where the expression of $\Omega$ is not given here for the sake of brevity. In view of \eqref{eq:gamma0sol}, we may regard $a$ as an implicit function $a=a(\lambda,\gamma)$ of $\lambda$ and $\gamma$ that satisfies \eqref{eq:gamma0sol}. Then \eqref{eq:dd} leads to a relation among $k$, $\lambda$ and $\gamma$, from which one can easily determine the variation of the wave number $k$ with respect to $\lambda$ and $\gamma$.

\subsection{Plateau--Rayleigh instability}

\begin{figure}[htbp!]
	\centering
	\subfloat[]{\includegraphics[width=0.4\textwidth]{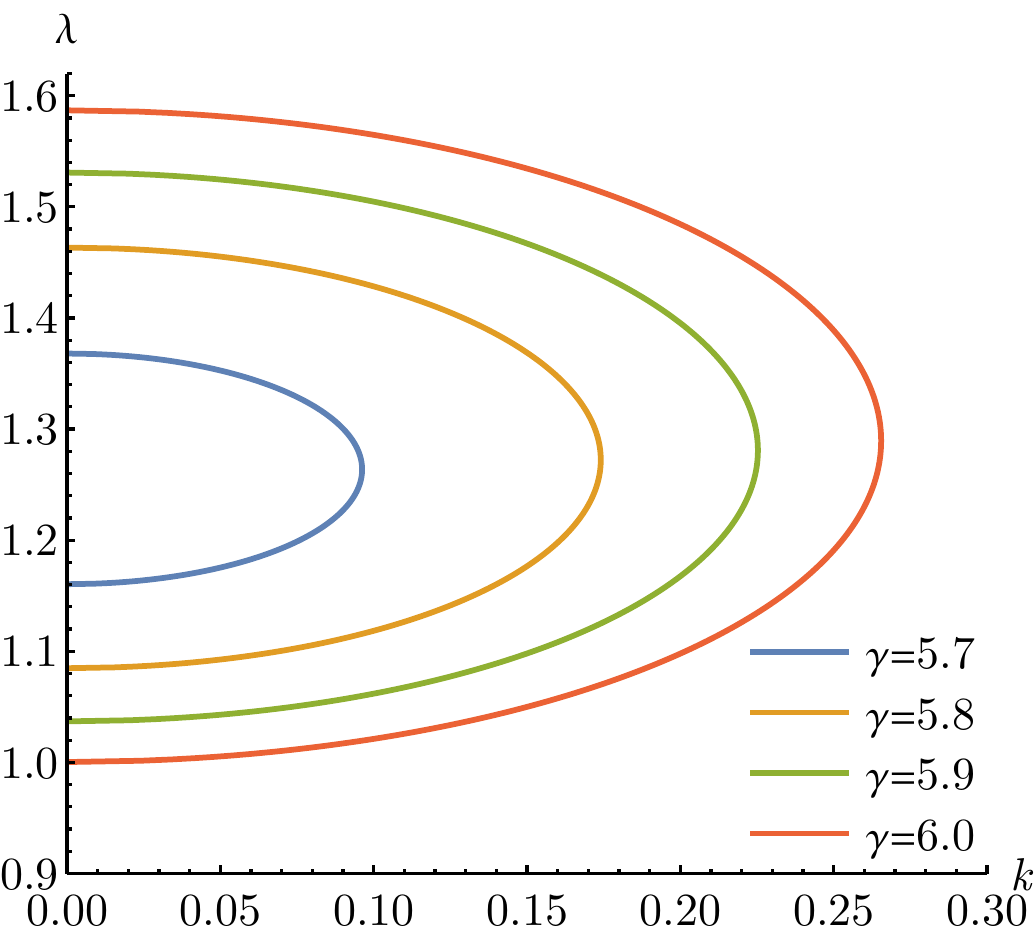}\label{subfig:c-a}
	}\qquad\qquad\qquad
	\subfloat[]{\includegraphics[width=0.4\textwidth]{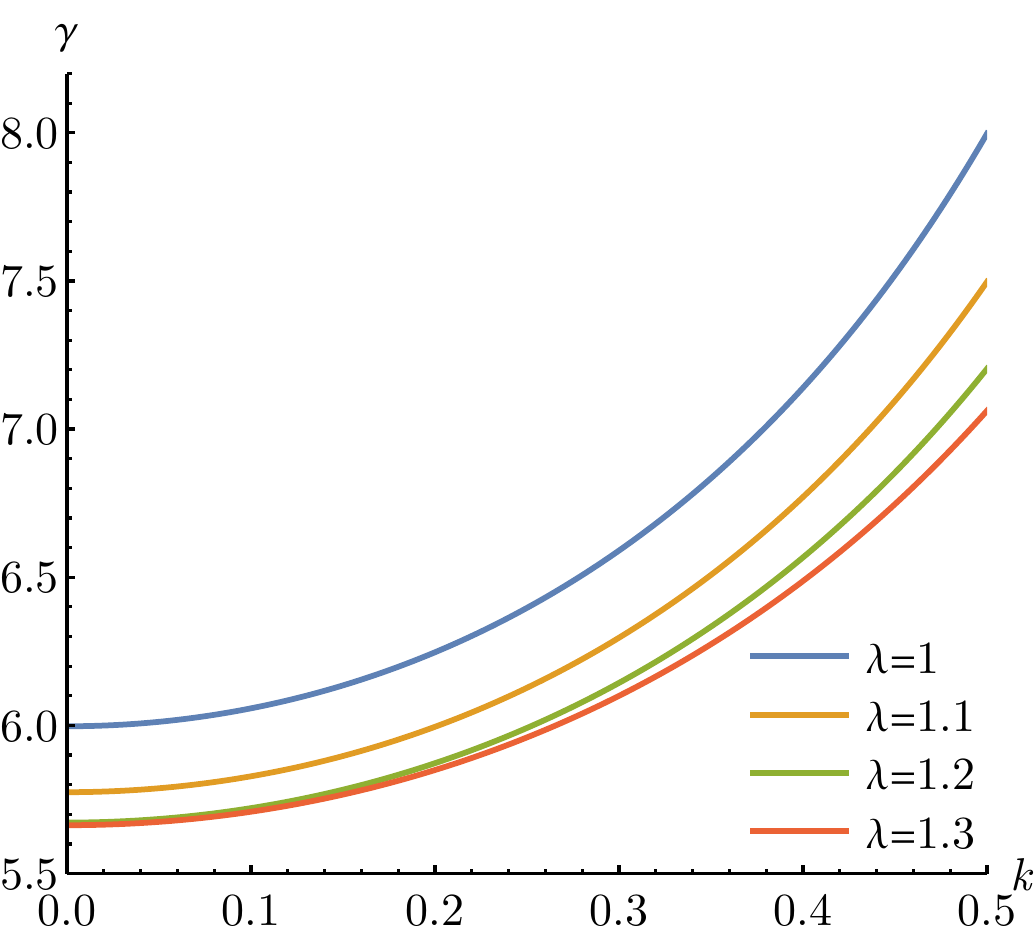}\label{subfig:c-b}
	}
	\caption{Variations of (a) $\lambda$ against $k$ at different fixed values of $\gamma$  and (b)  $\gamma$ against $k$ at different fixed values of $\lambda$. The cylinder is incompressible without surface stiffening, corresponding to the surface stiffness parameter  $\mu_s=0$ and $\lambda_s=0$.}
	\label{fig:c}
\end{figure}

We first validate the bifurcation condition \eqref{eq:dd} by comparing its predictions with the corresponding curves  in \cite{taffetani2015beading}. Setting $\nu= 0.5$, $\mu_s=0$ and $\lambda_s=0$, we depict in Fig. \ref{fig:c} the solutions of \eqref{eq:dd} for two typical loading scenarios where either $\gamma$ or $\lambda$ is fixed, showing perfect agreement with the corresponding curves in Fig. 1 of \cite{taffetani2015beading}.

\begin{figure}[htbp!]
	\centering
	\subfloat[]{\includegraphics[width=0.4\textwidth]{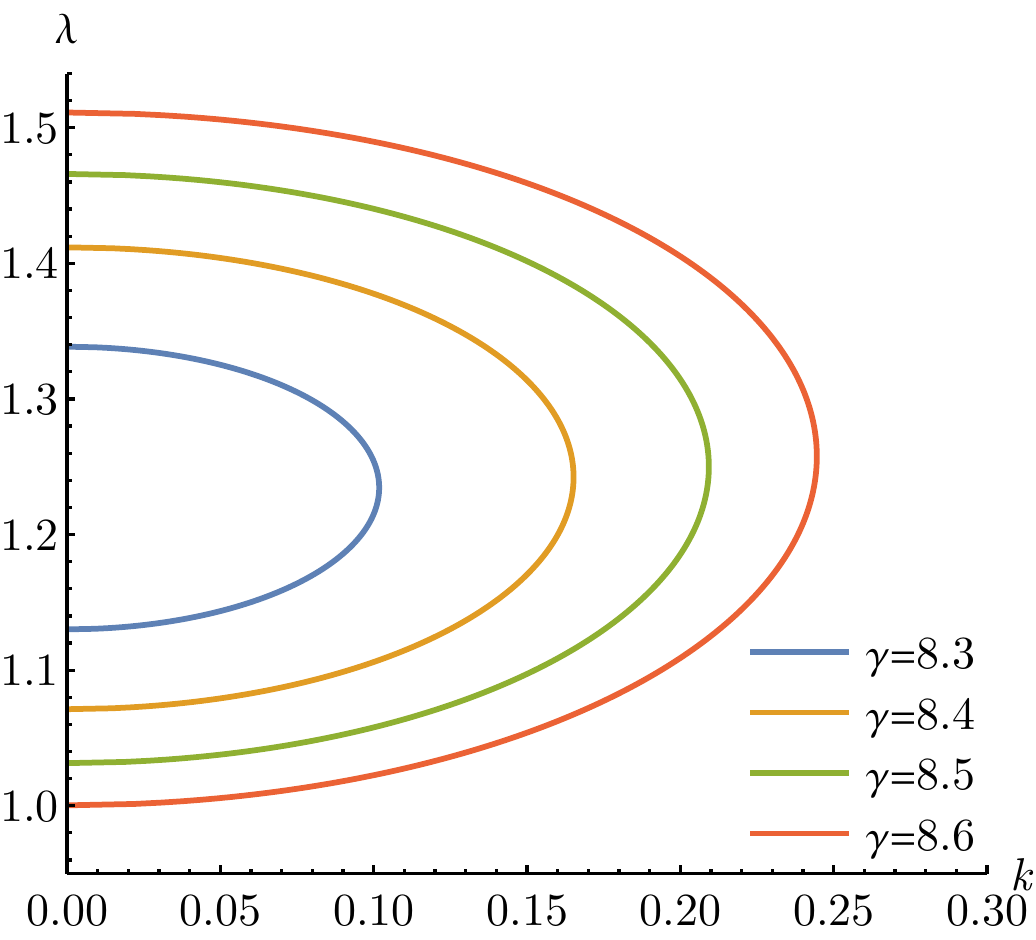}\label{subfig:cc-a}
	}\qquad\qquad\qquad
	\subfloat[]{\includegraphics[width=0.4\textwidth]{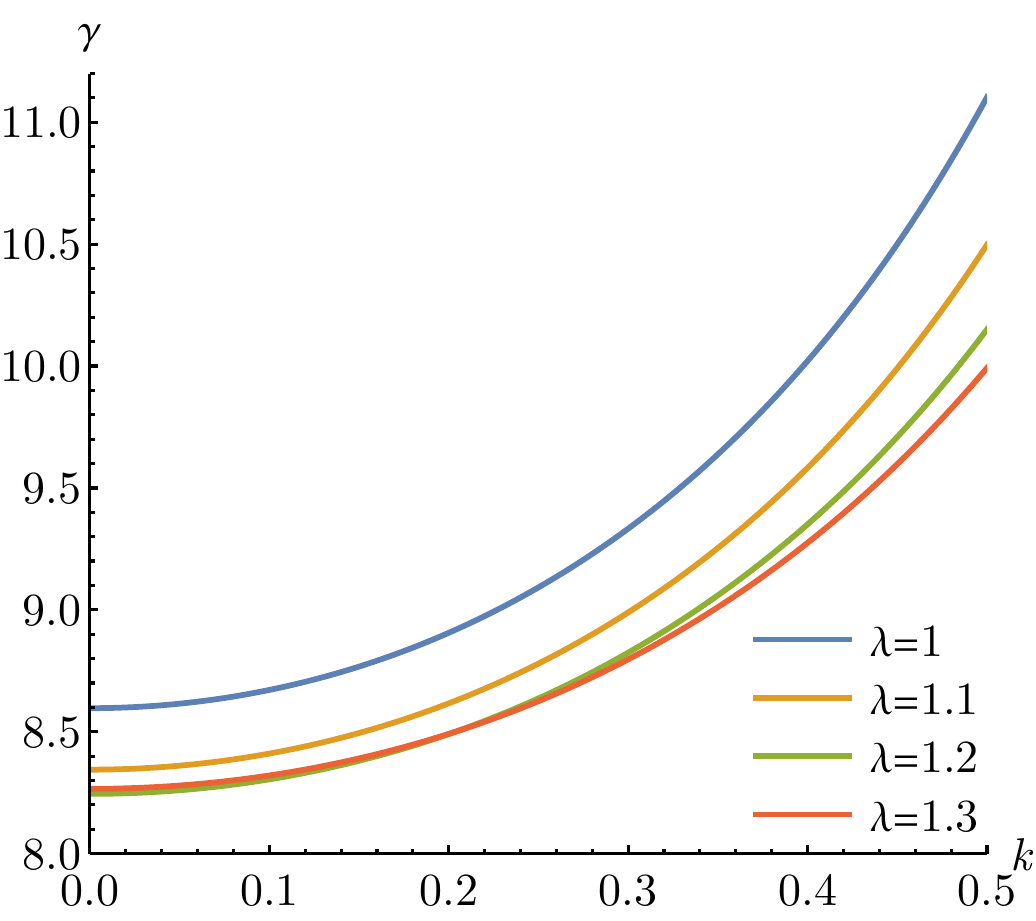}\label{subfig:cc-b}
	}
	\caption{Variations of (a) $\lambda$ against $k$ at different fixed values of $\gamma$  and (b)  $\gamma$ relative to $k$ at different fixed values of $\lambda$. The cylinder is incompressible and  the surface stiffness parameters are $\mu_s=0.2$ and $\lambda_s=0.6$.}
	\label{fig:cc}
\end{figure}

\begin{figure}[htbp!]
	\centering
	\subfloat[]{\includegraphics[width=0.4\textwidth]{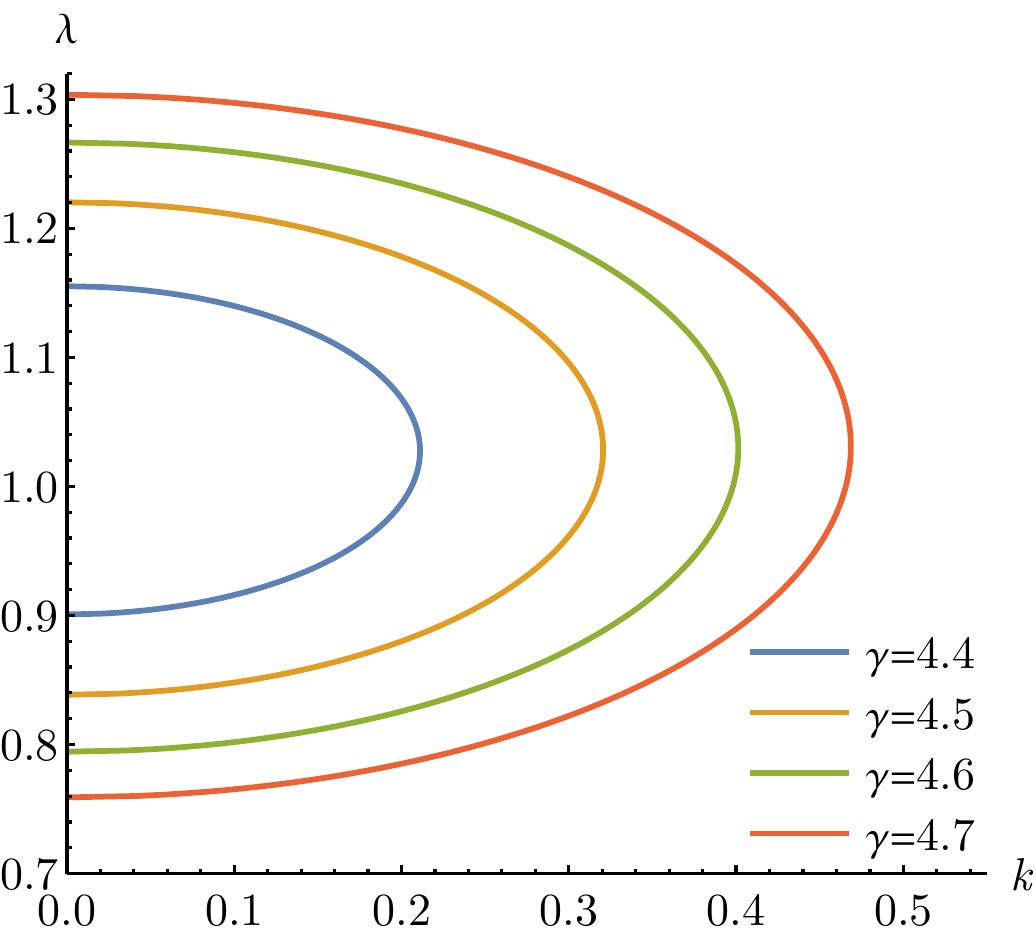}\label{subfig:d-a}
	}\qquad\qquad\qquad
	\subfloat[]{\includegraphics[width=0.4\textwidth]{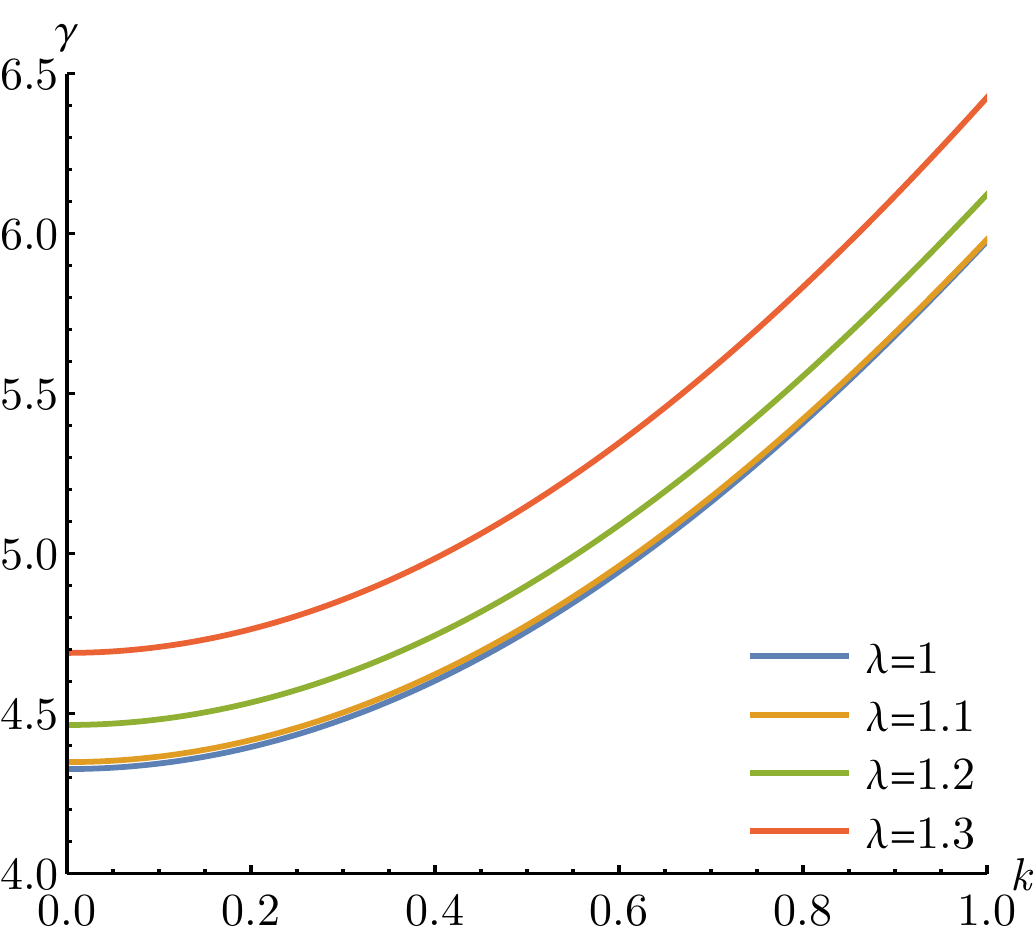}\label{subfig:d-b}
	}
	\caption{Variation of (a) $\lambda$ against $k$ at different fixed values of $\gamma$ and (b)  $\gamma$ against $k$ at different fixed values of $\lambda$. The cylinder is compressible with Poisson's ratio $\nu=0.4$ and no surface stiffening, corresponding to the surface stiffness parameter $\mu_s=0$ and $\lambda_s=0$.}
	\label{fig:d}
\end{figure}

\begin{figure}[htbp!]
	\centering
	\subfloat[]{\includegraphics[width=0.4\textwidth]{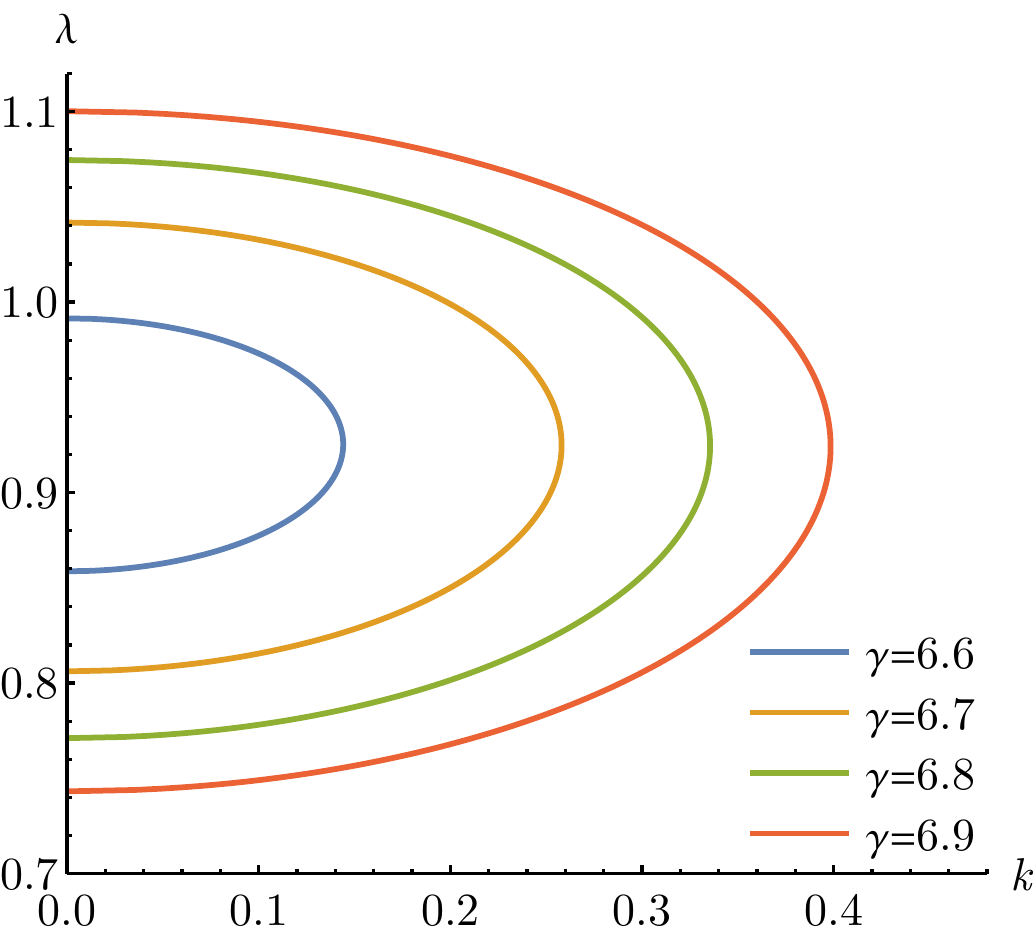}\label{subfig:dd-a}
	}\qquad\qquad\qquad
	\subfloat[]{\includegraphics[width=0.4\textwidth]{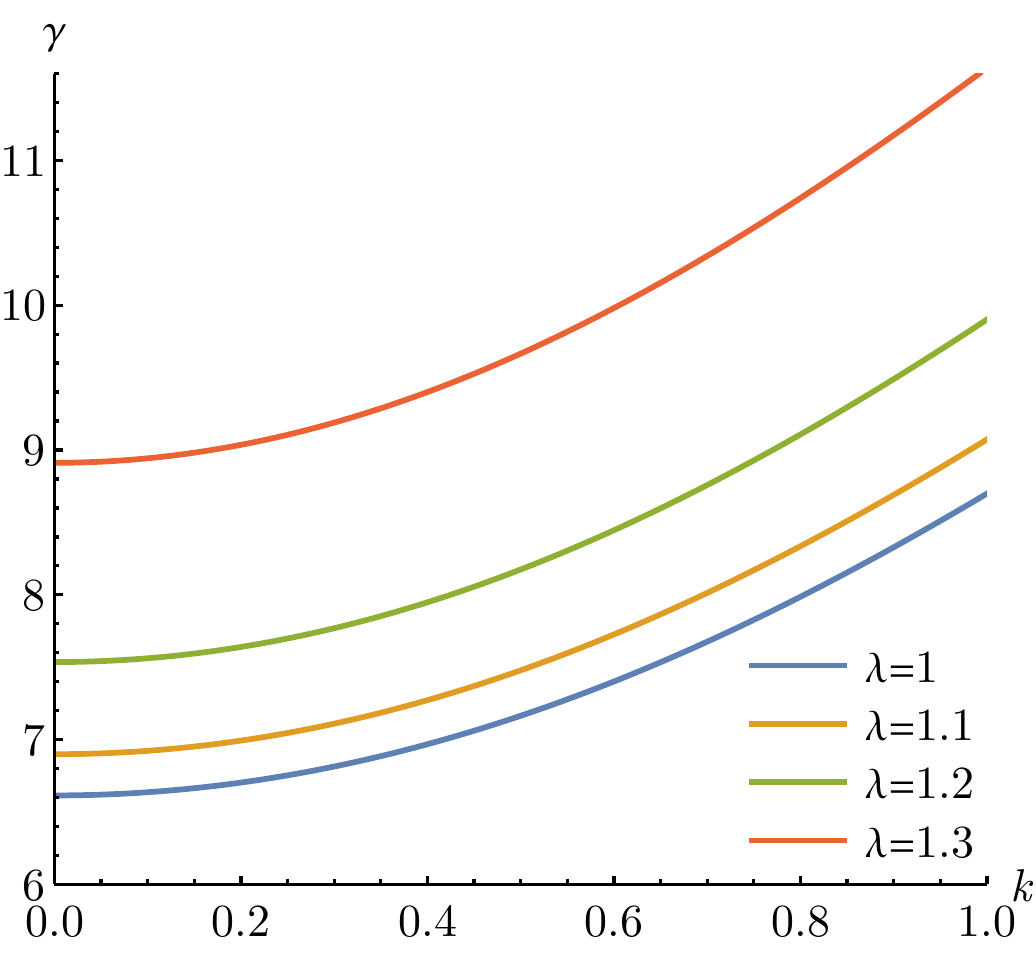}\label{subfig:dd-b}
	}
	\caption{Variations of (a) $\lambda$ against $k$ at different fixed values of $\gamma$  and (b)  $\gamma$ relative to $k$ at different fixed values of $\lambda$. The cylinder is compressible with Poisson's ratio $\nu=0.4$ and the surface stiffness parameters are $\mu_s=0.2$ and  $\lambda_s=0.6$.}
	\label{fig:dd}
\end{figure}

Next we consider the case when the surface stress is strain-dependent. The strain dependence or surface stiffening in other words is accounted for by allowing Lame's constants of the surface to be nonzero. For the representative parameters values $\nu=0.5$, $\mu_s=0.2$ and $\lambda_s=0.6$ which correspond to the surface Poisson ratio  $\nu_s=\lambda_s/(\lambda_s+2\mu_s)=0.6$, Fig. \ref{fig:cc} shows the variation of $\lambda$ or $\gamma$  with respect to $k$ when the other is kept fixed. The corresponding results for a compressible bulk solid with Poisson's ratio $\nu=0.4$ are presented in Figs. \ref{fig:d} and \ref{fig:dd}. It is seen that whether the bulk solid is incompressible or compressible, in both loading scenarios the smallest value of the load occurs at the wave number $k_\text{cr}=0$, which is the same as in the case of constant surface tension. We have verified this fact for a wide range of values of $\nu$, $\mu_s$ and $\lambda_s$. We thus conclude that the strain dependence is unlikely to affect the nature of the bifurcation and that the bifurcation  always takes place at a zero wave number. This is consistent with the findings in \cite{bakiler2023surface}.

Having established the fact that the bifurcation occurs at a zero wave number, we can determine the critical load analytically. Taking the limit $k\to 0$ in \eqref{eq:dd} and simplifying, we obtain
\begin{align}\label{eq:bif}
\begin{split}
&(2+\kappa+2\mu_s+\lambda_s+2a^2+2a^2\mu_s+a^2\lambda^2\lambda_s+3a^4\lambda^2\kappa)(2+\kappa+a^4\lambda^2\kappa-4a^2-2\lambda^2-4a^2\mu_s-4\lambda^2\mu_s)\\
&-4a^2(2+\kappa+2\mu_s+\lambda_s+a^2\lambda^2\lambda_s+a^4\lambda^2\kappa-2a^2-2a^2\mu_s)(2+2\mu_s+a^2\lambda^2\kappa)=0.
\end{split}
\end{align}
It is straightforward to verify that the above bifurcation condition is equivalent to
\begin{align}
J(\gamma,N)=\frac{\partial\gamma}{\partial a}\frac{\partial N}{\partial\lambda}-\frac{\partial\gamma}{\partial \lambda}\frac{\partial N}{\partial a}=0
\end{align}
with $\gamma$ and $N$ given by \eqref{eq:gamma0sol} and \eqref{eq:Nsol}, respectively. This equivalence is not a coincidence and can be established analytically; see \cite{yu2022analytic, yu2023one}.

When the bulk solid is incompressible for which $\kappa \to\infty$, we have $a=\lambda^{-1/2}$ and \eqref{eq:bif} reduces to
\begin{align}\label{eq:gamma0cr}
\gamma_{\text{cr}}=\frac{2(2+\lambda^3)}{\lambda^{3/2}}+\frac{2(2+\lambda+2\lambda^3)}{\lambda^{3/2}}\mu_s+\frac{\lambda_s}{\sqrt{\lambda}}.
\end{align}
The first term on the right-hand side represents the value of $\gamma_{\text{cr}}$ when $\mu_s=0$ and $\lambda_s=0$, and recovers the result  given by \cite{taffetani2015beading}. From \eqref{eq:gamma0cr}, it is clear that increasing $\mu_s$ or $\lambda_s$ leads to a rise of $\gamma_{\text{cr}}$, thus having a stabilizing effect on the Plateau--Rayleigh instability. It also follows from \eqref{eq:gamma0cr} that $\gamma_{\text{cr}}$ as a function of $\lambda$ has a minimum at $\lambda=\lambda_{\text{min}}$, which is characterized by the equation
\begin{align}
6(1+2\mu_s)\lambda_\text{min}^3-(2\mu_s+\lambda_s)\lambda_\text{min} -12(1+\mu_s)=0.
\end{align}
The unique real solution of the above equation for $\lambda_\text{min}$ is a monotonically decreasing function of $\mu_s$, equal to $2^{1/3}$ when
$\mu_s=0$ and tending to $1.055$ when $\mu_s \to \infty$. This stretch minimum has a special meaning in the post-bifurcation behavior. It was shown in \cite{fu2021necking} that if a solid cylinder is loaded by increasing $\gamma$ at a fixed length (i.e., fixed axial stretch), then when the critical value of $\gamma$ is reached, the bifurcation corresponds to localized necking if $\lambda<\lambda_\text{min}$ and to localized bulging if $\lambda>\lambda_\text{min}$, both cases being subcritical. In the exceptional case when $\lambda=\lambda_\text{min}$, the bifurcation is supercritical and the cylinder evolves smoothly into a \lq\lq two-phase" state. Although these results were obtained for the case when $\mu_s=0$ and $\lambda_s=0$, they are expected to be also valid when $\mu_s$ and $\lambda_s$ are non-zero.

\begin{figure}[htpb!]
	\centering
	\subfloat[]{\includegraphics[width=0.4\textwidth]{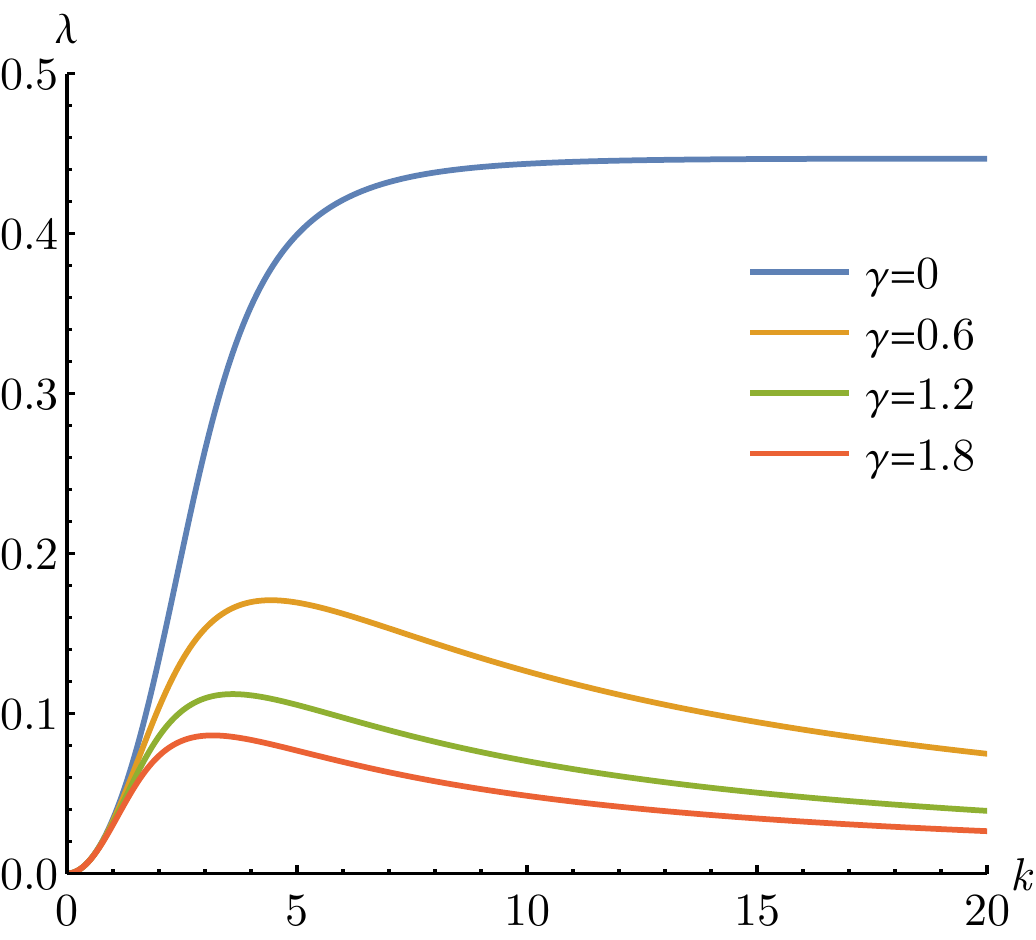}
	}\label{subfig:z-a}\qquad\qquad
	\subfloat[]{\includegraphics[width=0.4\textwidth]{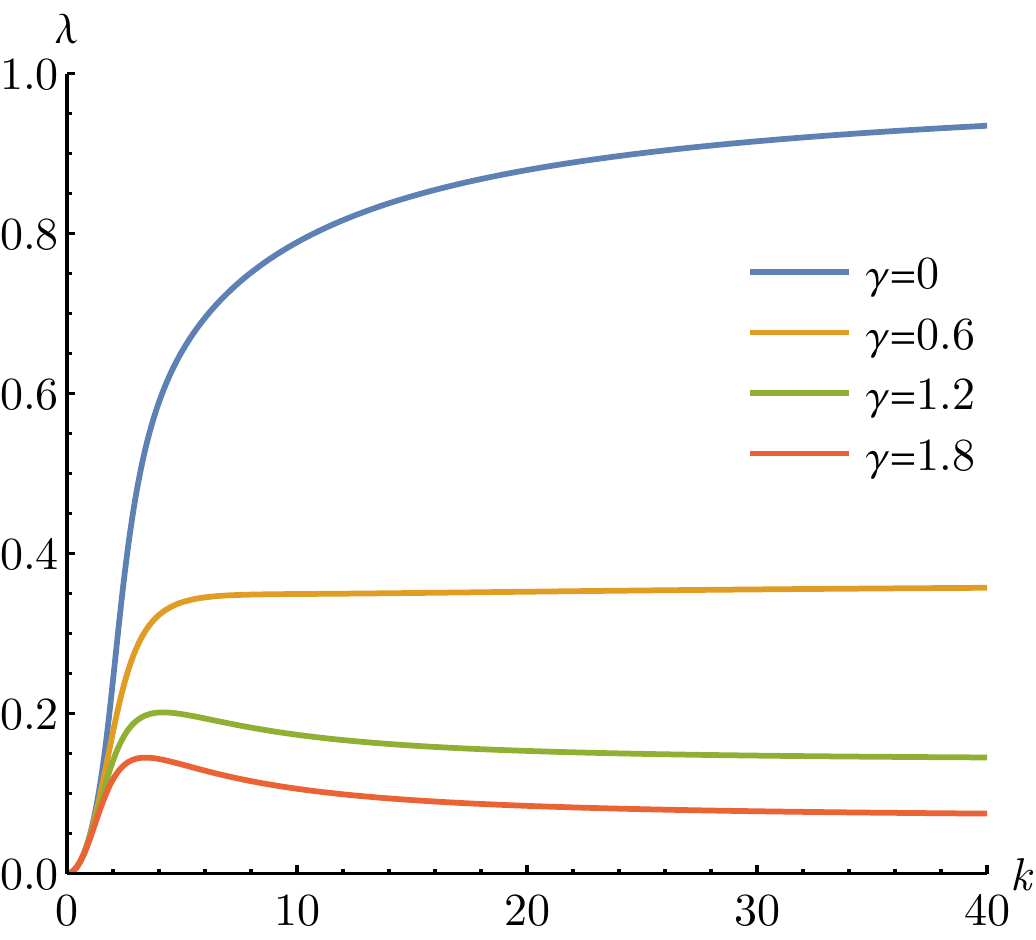}}\label{subfig:z-b}
	\caption{Variation of $\lambda$ against  $k$ at different fixed values of $\gamma$. The cylinder is incompressible subjected to a compressive axial force and the surface stiffness parameters are (a)  $\mu_s=0$, $\lambda_s=0$ and (b) $\mu_s=0.2$, $\lambda_s=0.6$.}
	\label{fig:z}
\end{figure}

\begin{figure}[htpb!]
	\centering
	\subfloat[]{\includegraphics[width=0.4\textwidth]{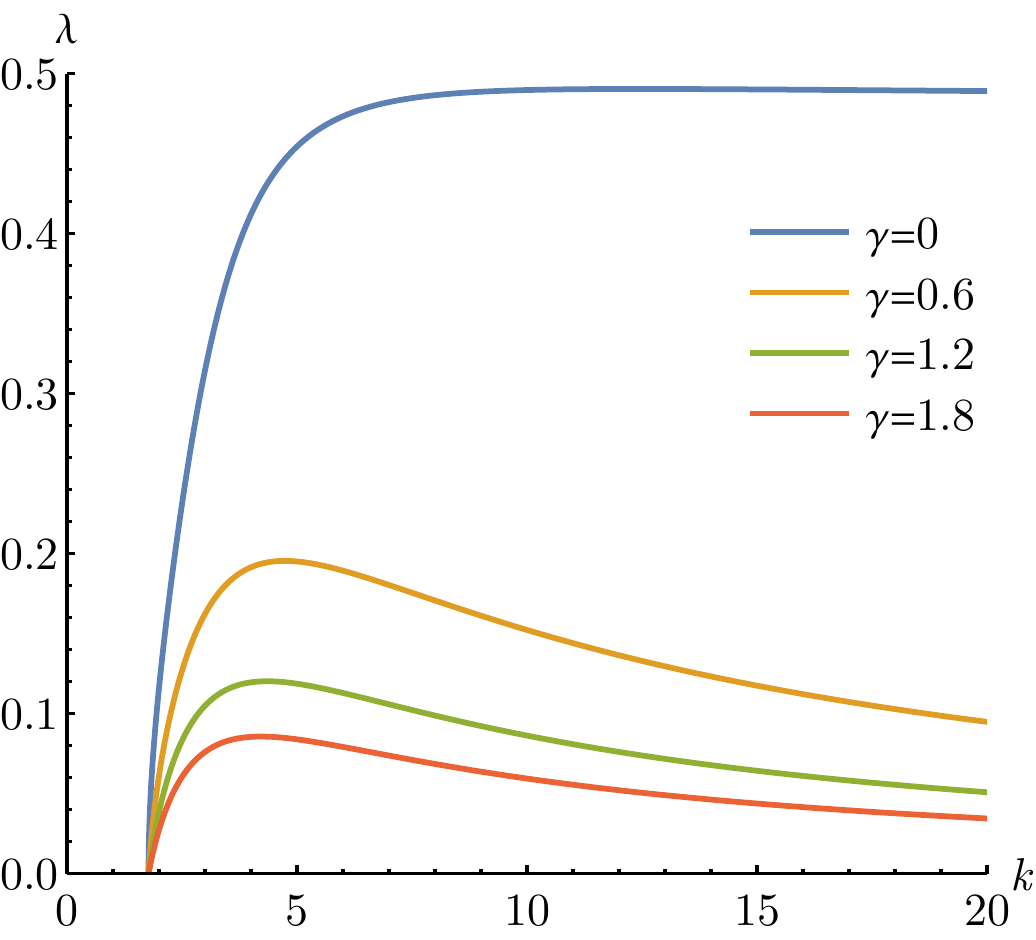}
	}\label{subfig:zz-a}\qquad\qquad
	\subfloat[]{\includegraphics[width=0.4\textwidth]{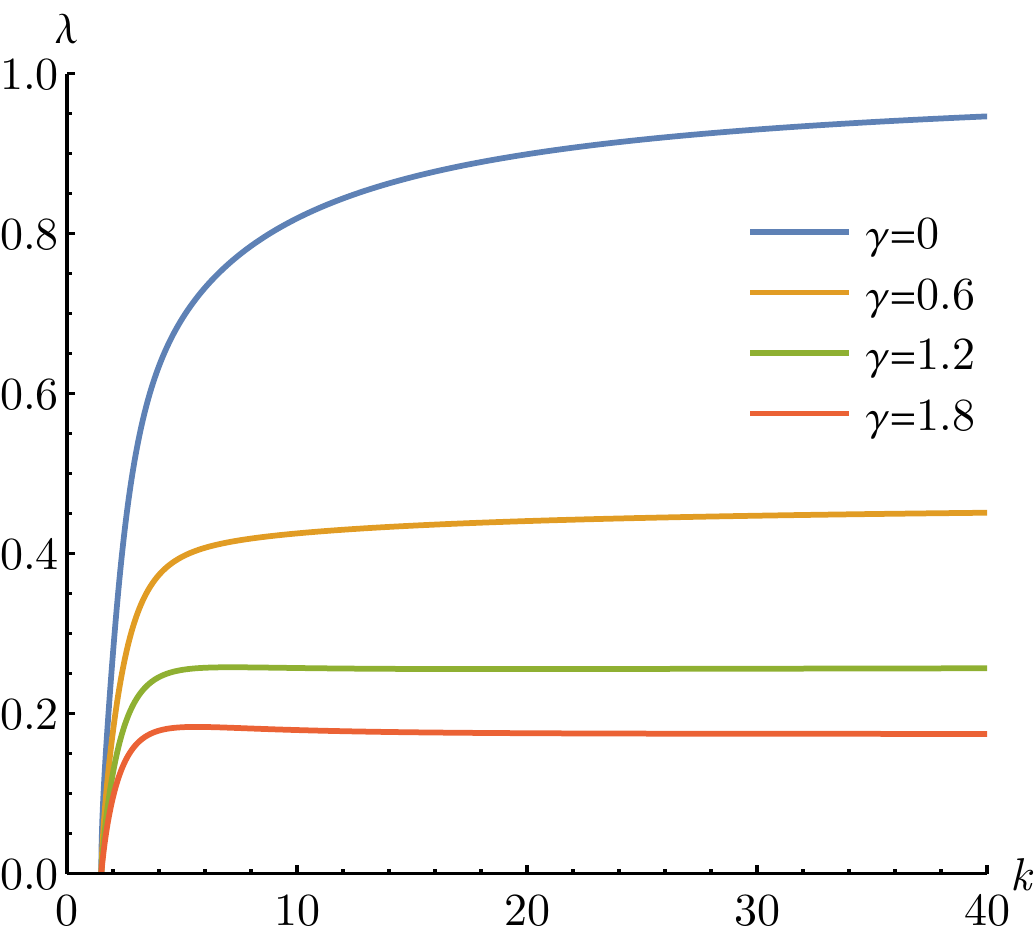}}\label{subfig:zz-b}
	\caption{Variation of $\lambda$ against  $k$ at different fixed values of $\gamma$. The cylinder is compressible with Poisson's ratio $\nu=0.4$ subjected to a compressive axial force and the surface stiffness parameters are  (a)  $\mu_s=0$, $\lambda_s=0$ and (b) $\mu_s=0.2$, $\lambda_s=0.6$.}
	\label{fig:zz}
\end{figure}

\subsection{Wilkes instability}
Finally, we consider the case when the cylinder is subjected to a compressive axial force as well as surface tension. Fig. \ref{fig:z} shows the variation of $\lambda$ with respect to $k$ for an incompressible cylinder without or with surface  stiffening. The corresponding results for a compressible cylinder with Poisson's ratio $\nu=0.4$ are given in Fig. \ref{fig:zz}. It is seen from Fig. \ref{fig:z}(a) that when $\gamma=0$ the maximum of $\lambda$ is $0.444$ and
is attained as $k$ goes
to infinity. This is consistent with the classical result of Wilkes \cite{wilkes1955stability}. The bifurcation mode is  essentially a surface wave mode localized near the surface of the cylinder.

From Figs. \ref{fig:z} and \ref{fig:zz}, we may draw the following conclusions. Firstly, results for a compressible cylinder are qualitatively similar to those for an incompressible cylinder. Secondly, when there is no surface stiffening, the maximum of $\lambda$ is attained at a finite wavenumber as soon as surface tension is present. For the situation with surface stiffening, the same conclusion holds as long as the surface tension is beyond a certain threshold (which is equal to $0.616$ and $1.039$ in Figs. \ref{fig:z}(b) and \ref{fig:zz}(b), respectively).  This means that surface tension penalizes the formation of small wavelength modes. Thirdly, in contrast with the case of Plateau--Rayleigh instability, surface stiffness tends to increase the maximum of $\lambda$ and thus has a destabilizing effect on the Wilkes instability.

\begin{figure}[htpb!]
	\centering
	\subfloat[]{\includegraphics[width=0.4\textwidth]{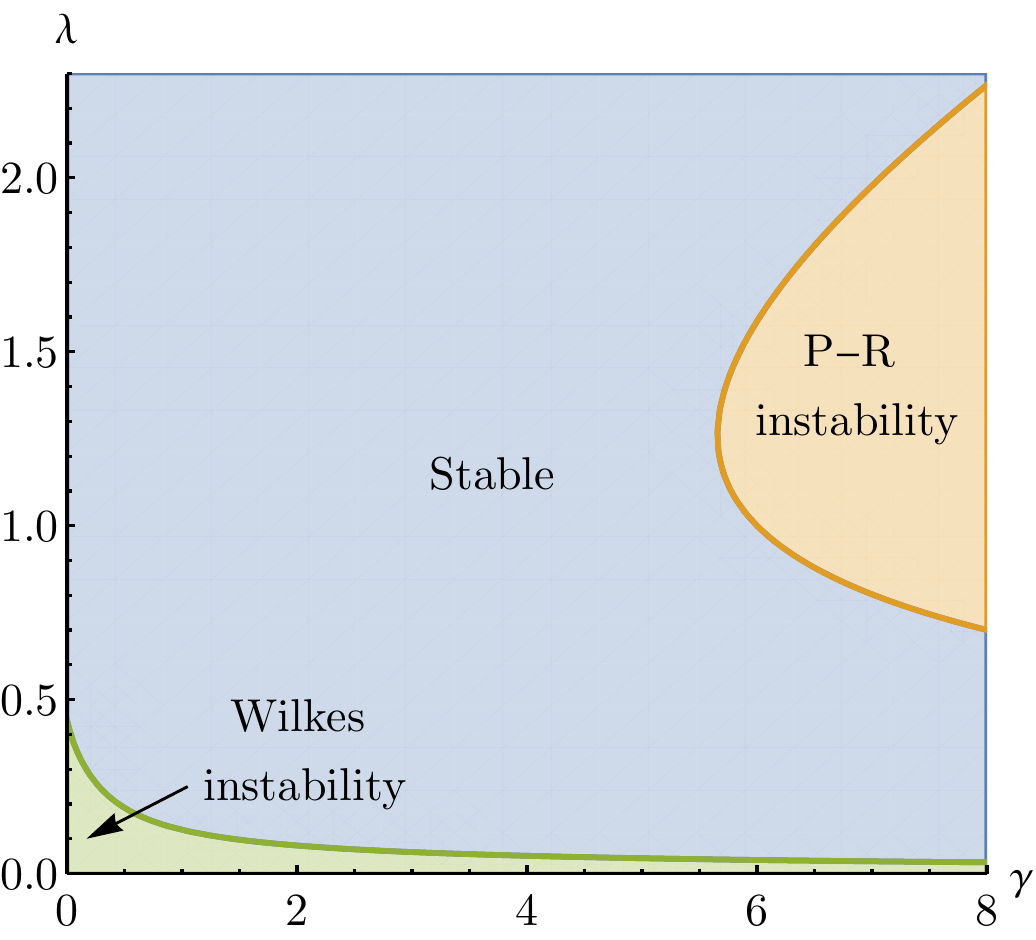}
	}\label{subfig:zw-a}\qquad\qquad
	\subfloat[]{\includegraphics[width=0.4\textwidth]{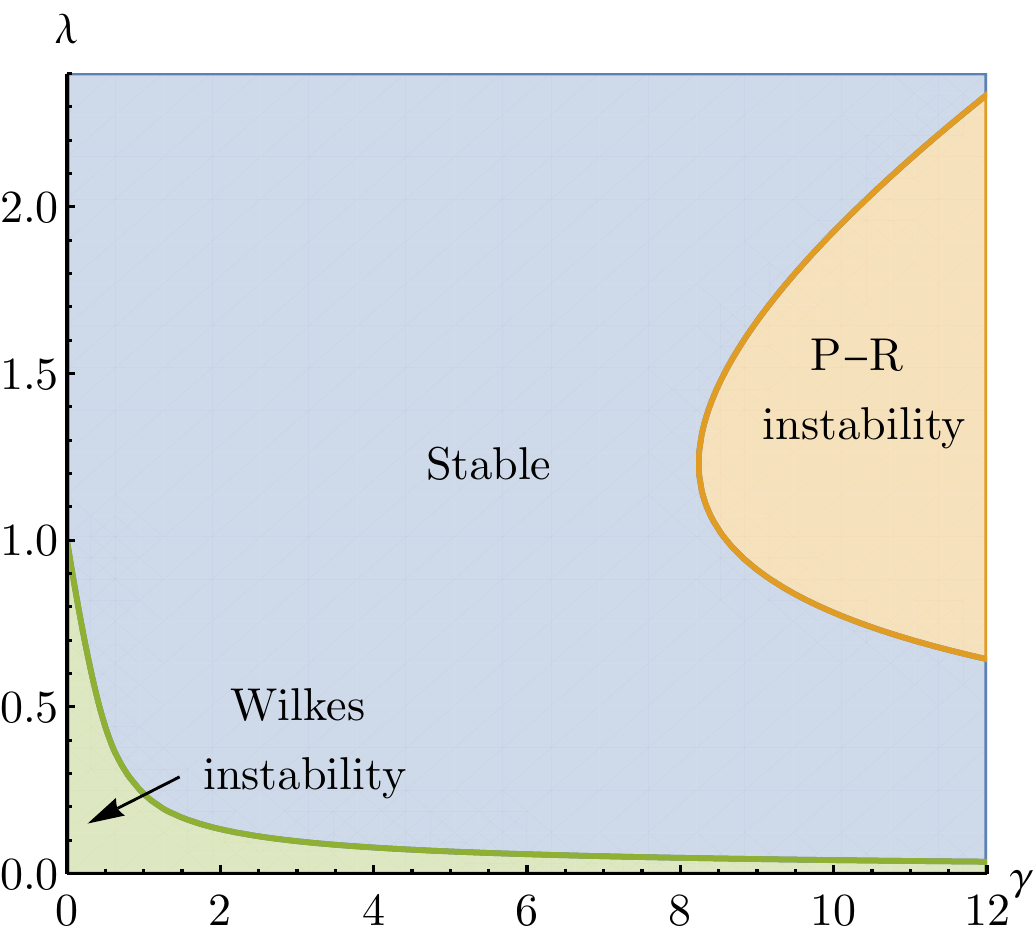}}\label{subfig:zw-b}
	\caption{Stability map for an incompressible cylinder as a function of axial stretch $\lambda$ and surface tension $\gamma$.  The surface stiffness parameters are (a)  $\mu_s=0$, $\lambda_s=0$ and (b) $\mu_s=0.2$, $\lambda_s=0.6$.}
	\label{fig:zw}
\end{figure}

\begin{figure}[htpb!]
	\centering
	\subfloat[]{\includegraphics[width=0.4\textwidth]{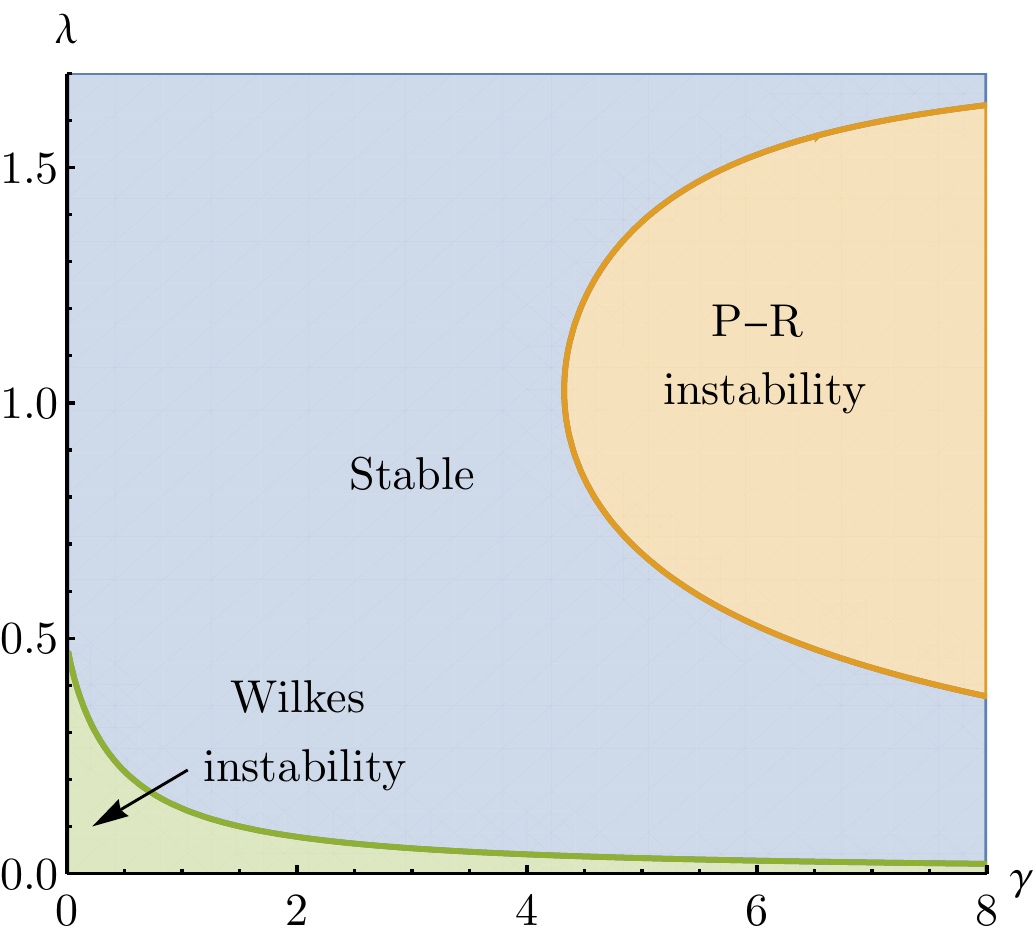}
	}\label{subfig:zx-a}\qquad\qquad
	\subfloat[]{\includegraphics[width=0.4\textwidth]{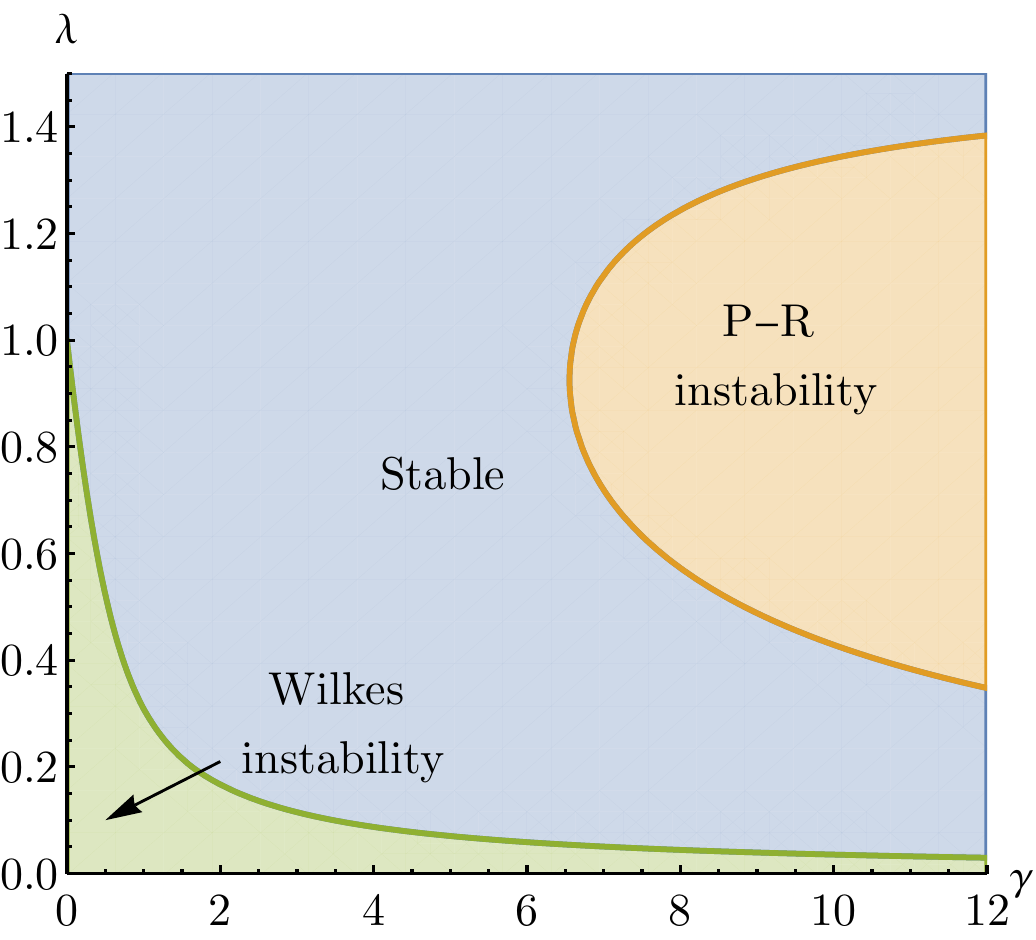}}\label{subfig:zx-b}
	\caption{Stability map for a compressible cylinder with Poisson's ratio $\nu=0.4$ as a function of axial stretch $\lambda$ and surface tension $\gamma$. The surface stiffness parameters are (a)  $\mu_s=0$, $\lambda_s=0$ and (b) $\mu_s=0.2$, $\lambda_s=0.6$.}
	\label{fig:zx}
\end{figure}

In Figs. \ref{fig:zw} and \ref{fig:zx} we display the bifurcation conditions for the Plateau--Rayleigh and Wilkes instabilities together. They provide a roadmap on how each instability arises when the cylinder is loaded in different ways. Similar stability maps were given by \cite{wang2020axisymmetric} for a hollow tube;  see, however, comments made by Emery and Fu \cite{emery2021localised}.
Finally, we remark that the bifurcation condition only provides a necessary condition in each case; what actually happens when the bifurcation condition is satisfied can only be determined by weakly and fully nonlinear analyses or numerical simulations.

\section{Conclusion}
Even in the simplest case when surface tension is a constant, the traction boundary condition involves the mean curvature of the surface (see \rr{kappa}), and it would not be a simple task to derive the incremental form of this curvature term, at least for a general surface. This may explain why recent studies on surface tension-induced instability have resorted to other alternative ways to derive the incremental boundary equations. On the other hand, it is clearly desirable to have access to incremental boundary conditions as simple as in the purely mechanical case so that their use does not require extra knowledge of differential geometry. It is precisely this consideration that has motivated our current study. Our main results are \rr{eq:chi3}, \rr{eq:chis},  and \rr{eq:calC} which are valid for any material and any geometry, and \rr{eq:C1} which is valid for an isotropic surface. Our approach is obviously inspired by Professor Ray Ogden's style of presentation, and so we feel that it is appropriate to dedicate this work to Professor Ogden on the occasion of his 80th birthday.

We conclude by remarking that we have only considered surface elasticity whereby the surface tension is a function of the surface deformation gradient. A higher-order surface elasticity theory that allows the surface tension to depend on the gradient of the surface deformation gradient so as to capture the bending stiffness of the surface has been developed by Steigmann and Ogden \cite{steigmann1999elastic}, and Gao et al. \cite{gao2014curvature} among others. A parallel derivation of the associated incremental boundary conditions may be carried out although they will be considerably more involved.

\begin{acks}
This work was supported by the National Natural Science Foundation of China (Grant No 12072224) and the Engineering and Physical Sciences Research Council, UK (Grant No EP/W007150/1).
\end{acks}

\renewcommand{\theequation}{A.\arabic{equation}}
\setcounter{equation}{0}  

\section*{Appendix A Simple expressions of surface moduli for isotropic materials}\label{app:A}

In this appendix, we give a derivation of the simple expressions of surface  moduli announced in Subsection \ref{sec:surface}.

In view of the assumption that the unit basis vectors $\{\bm{e}_1, \bm{e}_2\}$ coincide with the principal axes of stretch on the finitely deformed surface and $\bm{e}_3$ is the unit normal to this surface, $\bar{\bm{F}}_s$ has the diagonal matrix representation $\bar{\bm{F}}_s=\diag(\lambda_1,\lambda_2,0)$ relative to the ordered basis $(\bm{e}_1,\bm{e}_2,\bm{e}_3)$. Let $\bar{\bm{F}}_s$ be perturbed by a small superficial tensor $\bm{\varepsilon}$,
\begin{align}
\bm{F}_s=\bar{\bm{F}}_s+\bm{\varepsilon}=\begin{pmatrix}
\lambda_1+\varepsilon_{11} & \varepsilon_{12} & 0\\
\varepsilon_{21} & \lambda_2 +\varepsilon_{22} & 0\\
\varepsilon_{31} & \varepsilon_{32} & 0
\end{pmatrix}.
\end{align}
To the second order in terms of $\bm{\varepsilon}$, the eigenvalues of $\sqrt{\bm{F}_s^T\bm{F}_s}$  are given by
\begin{align}\label{eq:lambdas}
\begin{split}
&\lambda^s_1=\lambda_1+\varepsilon_{11}+\frac{\lambda_1^2(\varepsilon_{12}^2+\varepsilon_{21}^2)+2\lambda_{1}\lambda_2\varepsilon_{12}\varepsilon_{21}+(\lambda_1^2-\lambda_2^2)\varepsilon_{31}^2}{2\lambda_1(\lambda_1^2-\lambda_2^2)}+O(|\bm{\varepsilon}|^3),\\
&\lambda^s_2=\lambda_2+\varepsilon_{22}+\frac{\lambda_2(\varepsilon_{12}^2+\varepsilon_{21}^2)+2\lambda_1\lambda_2\varepsilon_{12}\varepsilon_{21}+(\lambda_2^2-\lambda_1^2)\varepsilon_{32}^2}{2\lambda_2(\lambda_2^2-\lambda_1^2)}+O(|\bm{\varepsilon}|^3).
\end{split}
\end{align}
Substituting \eqref{eq:lambdas} into the surface energy function $\varGamma(\lambda_1^s,\lambda_2^s)$ and expanding the resulting expression  to second order, one  can express the derivatives $\frac{\partial^2 \varGamma}{\partial F^s_{ik }\partial F^s_{jl}}\Big|_{\bar{\bm{F}}_s}$ in terms of the derivatives of $\varGamma$ with respect to $\lambda_1^s$ and $\lambda_2^s$ by identifying the coefficients of $\varepsilon_{ik}\varepsilon_{jl}$. Inserting these into \eqref{eq:calC} then yields the surface moduli given in \eqref{eq:C1}. Alternatively, these expressions can be derived using the procedure employed by Ogden \cite{ogden1984non} to derive \rr{eqn3}.

\end{document}